\documentclass[]{interact}

\usepackage[T1]{fontenc}
\usepackage{lmodern}
\usepackage{epstopdf}
\usepackage{booktabs}
\usepackage{float}
\usepackage{placeins}
\usepackage[caption=false]{subfig}
\usepackage[numbers,sort&compress]{natbib}
\bibpunct[, ]{[}{]}{,}{n}{,}{,}
\renewcommand\bibfont{\fontsize{10}{12}\selectfont}
\makeatletter
\def\NAT@def@citea{\def\@citea{\NAT@separator}}
\makeatother
\usepackage[hidelinks]{hyperref}

\graphicspath{{figures/}}

\begin{document}


\title{Physical-Field Reconstruction from Sparse Observations: When Are Diffusion Models Preferable to Deterministic Regression?}
\author{
\name{Hao Zhou\textsuperscript{a}, Rui Zhang\textsuperscript{a}, Qi Wang\textsuperscript{b} and
Hao Sun\textsuperscript{a}\thanks{CONTACT Rui Zhang and Hao Sun. Email: \{rayzhang,haosun\}@ruc.edu.cn}}
\affil{\textsuperscript{a}Gaoling School of Artificial Intelligence, Renmin University of China, Beijing, China\\
\textsuperscript{b}Zhongguancun Academy, Beijing, China}
}
\history{Compiled September 5, 2026}

\maketitle

\begin{abstract}
Reconstructing physical fields from sparse observations is central to system identification, forecasting, and control, yet sparse measurements generally underdetermine the full field. This makes reconstruction an ill-posed inverse problem rather than simple interpolation. Although many deterministic and generative methods have been developed, there is still no clear consensus on when a single point estimate is sufficient and when a distribution of plausible reconstructions is more useful. We conduct a fair comparison of a deterministic U-Net, conditional diffusion, and prior-guided diffusion under matched experimental settings, including 2D Poisson equation, 2D Navier--Stokes flow, and 1D Kuramoto--Sivashinsky dynamics. Through this comparison, we make three observations. First, accuracy is field- and regime-dependent, with no systematic advantage for diffusion under higher complexity or sparser observations. Second, ensemble means improve phase-aligned accuracy, whereas individual samples better preserve variability and can retain high-wavenumber power in selected regimes. Third, conditional diffusion provides more reliable uncertainty estimates at lower cost, while prior-guided diffusion is more robust to mask-distribution shifts but requires substantially higher inference cost and guidance tuning. These results clarify when generative reconstruction is useful and provide guidance for improving uncertainty estimation, fine-scale sample fidelity, robustness, and computational efficiency in sparse field reconstruction.
\end{abstract}


\section{Introduction}
\label{sec:introduction}

Reconstructing physical fields from sparse observations is a fundamental problem in state estimation, system identification, forecasting, and control across science and engineering~\cite{brunton2020machine,lu2022partial,course2023state}. 
The problem is generally ill posed because multiple fields can agree with all available measurements while differing at unobserved locations. This underdetermination makes sparse reconstruction more than an interpolation problem: the observations constrain only part of the field, while the remaining degrees of freedom must be inferred from additional statistical, geometric, or physical structure. Existing methods differ mainly in how they introduce such structure, such as reduced-order bases, learned reconstruction operators, and generative field priors.

Classical approaches to recovering the physical field have relied on reduced-order and sparse representations, including proper orthogonal decomposition (POD) and its gappy extensions~\cite{berkooz1993pod,everson1995gappy,willcox2006gappy} and compressed sensing~\cite{donoho2006compressed,callaham2019sparse}.
Deep learning-based approaches have further expanded this paradigm by exploiting the representation-learning capacity of neural networks to learn nonlinear reconstruction operators from data. This flexibility allows sparse observations and reconstructed fields to be represented in several ways.
For predefined sensor layouts, neural networks infer complete discretized fields either directly from sensor values or through reduced field representations~\cite{erichson2020shallow,nair2020leveraging,dubois2022flow}.
Grid-based models exploit spatial structure in coarse or incomplete fields using convolutional and operator-learning architectures~\cite{fukami2019superres,mo2025reconstructing,zhang2025energy}.
For irregular or variable sensor layouts, Voronoi-assisted networks convert scattered measurements into structured inputs, whereas attention-based architectures process sensor values and coordinates directly~\cite{fukami2021voronoi,santos2023senseiver}.
Coordinate-based implicit neural representations further allow reconstructed fields to be evaluated continuously at arbitrary locations~\cite{luo2024continuous,guoze2026geometryaware}.
When governing equations are available, physics-informed methods can further incorporate them through PDE-residual constraints~\cite{raissi2019physics,raissi2020hidden}.
These developments have improved the flexibility and expressive power of sparse field reconstruction. However, the resulting reconstructions are generally deterministic: for a given set of observations, they return a single field estimate and do not represent the distribution of fields compatible with the measurements.

Beyond deterministic approaches, Bayesian inverse methods address the non-uniqueness of sparse reconstruction by characterizing a posterior distribution over fields conditioned on the observations~\cite{stuart2010inverse}. 
Modern generative models provide flexible data-driven representations of field priors or conditional distributions and can produce multiple plausible reconstructions consistent with the same measurements. 
Variational autoencoders, generative adversarial networks, diffusion models, and flow matching~\cite{kingma2014vae,goodfellow2014gan,ho2020ddpm,song2021score,lipman2023flowmatching} have been adapted to physical-field generation and sparse reconstruction~\cite{gundersen2021semiconditional,guemes2022raseedgan,du2024confield,zhou2026perflow,oommen2026turbulence}. 
Within diffusion-based reconstruction, observations can condition the model during training or guide a learned full-field prior during sampling~\cite{shysheya2024conditional}. 
Conditional diffusion uses observations as conditioning inputs during training and directly learns an observation-conditioned field distribution~\cite{baldassari2023conditional,li2025palsb,oommen2026turbulence}.
Prior-guided diffusion instead learns an unconditional prior from complete fields and imposes measurement consistency during sampling through data-consistency corrections or gradient-based guidance~\cite{chung2023dps,shu2023physicsdiffusion,du2024confield,li2024s3gm,huang2024diffusionpde,amoros2026guiding}. These methods shift sparse reconstruction from predicting a single best estimate toward representing a family of observation-consistent fields, making them particularly attractive when ambiguity, uncertainty, or sample diversity is scientifically relevant.

Existing studies have demonstrated the feasibility of these reconstruction approaches, but differences in benchmarks, architectures, observation protocols, physical constraints, and sampling procedures limit direct comparison across formulations. Consequently, it remains unclear when deterministic regression is sufficient and when generative reconstruction provides practical advantages in accuracy, uncertainty quantification, robustness, or computational cost. We address this gap through a controlled comparison of a deterministic U-Net~\cite{ronneberger2015unet} and two diffusion formulations based on the same EDM framework~\cite{karras2022edm}, namely conditional EDM (C-EDM) and prior-guided EDM (G-EDM). Model backbones, data splits, and observation protocols are matched across methods.  The evaluation covers four reconstruction targets drawn from the Poisson equation, two-dimensional Navier--Stokes flow, and one-dimensional Kuramoto--Sivashinsky dynamics, with controlled variations in field complexity, observation density, and mask distribution. The principal findings are summarized below.
\begin{enumerate}
  \item Reconstruction accuracy is field-specific, and neither increasing field complexity nor reducing observation density systematically favors generative methods.

  \item Generative models produce ensembles of plausible reconstructions that support uncertainty quantification. Their ensemble means improve phase-aligned accuracy but can attenuate high-wavenumber power in selected regimes.

  \item Conditional diffusion generally provides more reliable uncertainty estimates at lower cost, whereas prior-guided diffusion is least sensitive to mask-distribution shifts but requires longer sampling and guidance-weight tuning.
\end{enumerate}  

In summary, these findings clarify the trade-offs among reconstruction accuracy,
spectral fidelity, uncertainty quantification, robustness, and computational cost,
providing a basis for selecting reconstruction strategies suited to different
scientific objectives and sensing conditions. Moreover, these observations suggest concrete ways to improve future sparse-reconstruction methods. For instance, conditional generative methods should better preserve fine-scale sample variability without degrading pointwise accuracy, while prior-guided methods should retain their flexibility under changing sensing geometries while reducing sampling cost and guidance sensitivity.

The remainder of this paper is organized as follows. Section~\ref{sec:methods}
formulates the sparse reconstruction problem and introduces the deterministic
U-Net, C-EDM, and G-EDM formulations. Section~\ref{sec:protocol} describes the
benchmark PDE systems, observation protocols, controlled comparison setup, and
evaluation metrics. Section~\ref{sec:results} presents the empirical comparison
in terms of reconstruction accuracy, spectral behavior, robustness to
mask-distribution shifts, uncertainty estimation, and inference cost.
Section~\ref{sec:discussion} discusses the implications and limitations of the
findings, and Section~\ref{sec:conclusion} concludes the paper.

\section{Methods}
\label{sec:methods}

\subsection{Problem formulation and reconstruction paradigms}
\label{sec:problem}

Let $\mathbf{x}_0\in\mathbb{R}^{C\times|\mathcal{G}|}$ denote a complete
field with $C$ channels on a spatial or spatiotemporal grid $\mathcal{G}$.
Let $\mathbf{M}\in\{0,1\}^{|\mathcal{G}|}$ denote the binary observation mask,
with $M_g=1$ at observed locations and $M_g=0$ elsewhere. The mask is shared
across channels when $C>1$. The corresponding zero-filled observation is
\begin{equation}
  \mathbf{y}=\mathbf{M}\odot\mathbf{x}_0,
  \label{eq:observation}
\end{equation}
where $\odot$ denotes elementwise multiplication. Sparse physical-field
reconstruction seeks to infer $\mathbf{x}_0$ from
$(\mathbf{y},\mathbf{M})$.

\begin{figure}[!t]
  \centering
  \includegraphics[width=\textwidth]{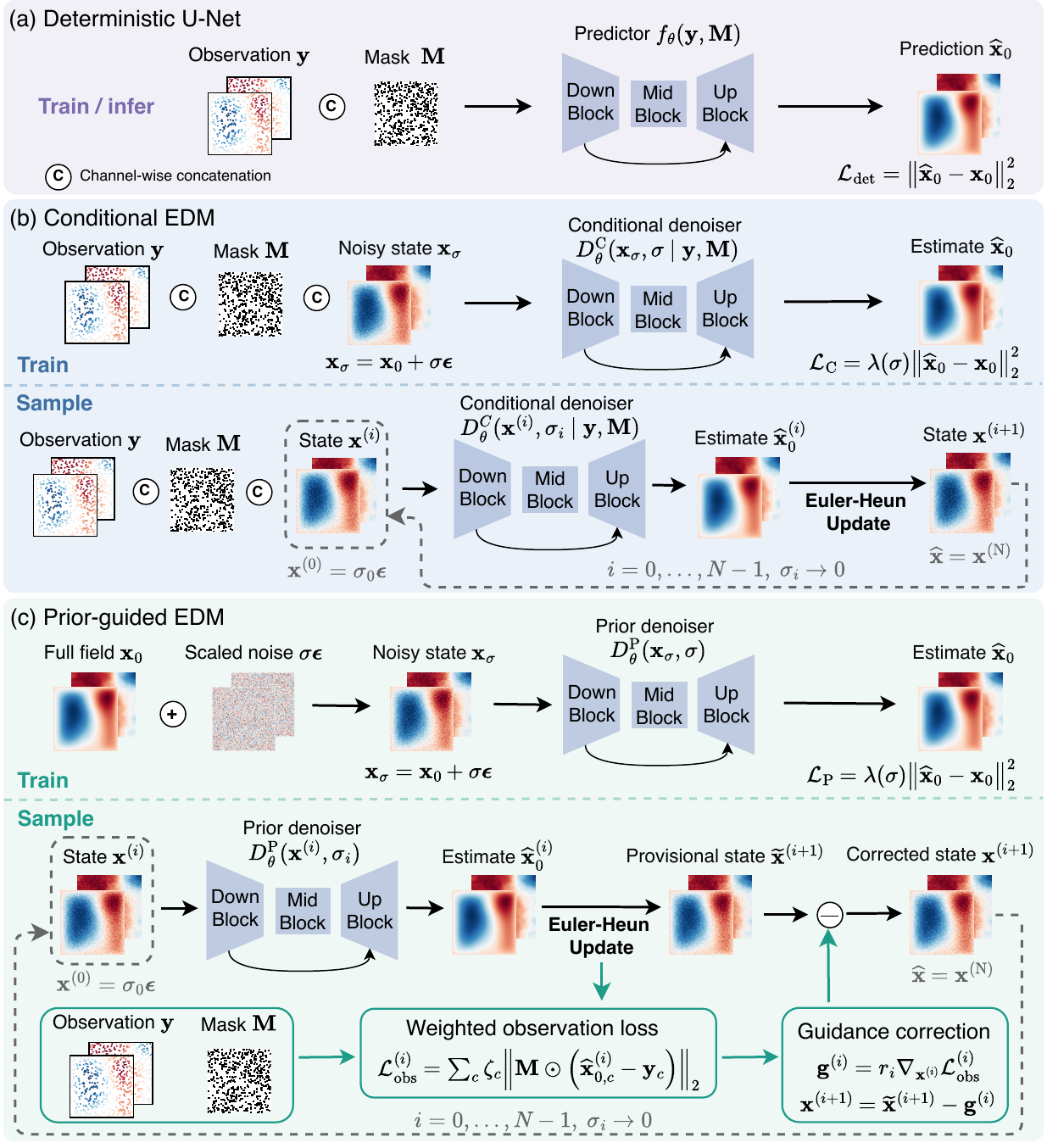}
  \caption{Three architecture-matched formulations for physical-field reconstruction from sparse observations. (a) U-Net maps the zero-filled observation and mask directly to a single reconstruction. (b) C-EDM conditions each denoising evaluation on the observation and mask. (c) G-EDM learns a full-field prior and incorporates the observations through gradient-based corrections during sampling.}
  \label{fig:method_overview}
\end{figure}

Sparse observations generally admit multiple full-field reconstructions consistent with the same measurements. 
We therefore compare three reconstruction formulations. The deterministic U-Net learns a direct map from
\((\mathbf{y},\mathbf{M})\) to a single reconstruction
\(\widehat{\mathbf{x}}_0\).
Conditional EDM (C-EDM) models
$p_\theta(\mathbf{x}_0\mid\mathbf{y},\mathbf{M})$ by conditioning the denoising process on the observations and mask.
Prior-guided EDM (G-EDM) instead models the full-field distribution $p_\theta(\mathbf{x}_0)$ and incorporates the observations through guidance during sampling.
Here, EDM refers to the diffusion formulation and denoiser preconditioning introduced by Karras et al.~\cite{karras2022edm}.
Figure~\ref{fig:method_overview} summarizes how observations enter the three formulations.

\subsection{Deterministic U-Net}
\label{sec:deterministic_method}

The deterministic formulation treats reconstruction as direct
regression. The U-Net~\cite{ronneberger2015unet} $f_{\theta}^{\mathrm{D}}$
receives the channel-wise concatenation of the zero-filled observations and
mask and predicts the complete field as 
\begin{equation}
  \widehat{\mathbf{x}}_0
  =
  f_{\theta}^{\mathrm{D}}
  \!\left(\operatorname{cat}[\mathbf{y},\mathbf{M}]\right).
  \label{eq:unet_prediction}
\end{equation}
The model is trained by minimizing the expected full-field mean-squared error
\begin{equation}
  \mathcal{L}_{\mathrm{D}}
  =
  \mathbb{E}_{\mathbf{x}_0,\mathbf{M}}
  \left[
    \operatorname{MSE}
    \!\left(\widehat{\mathbf{x}}_0,\mathbf{x}_0\right)
  \right],
  \label{eq:unet_loss}
\end{equation}
where the mean is taken over all channels and grid locations. At inference, a single forward pass produces one reconstruction.

\subsection{Conditional EDM (C-EDM)}
\label{sec:conditional_edm}

C-EDM models the conditional field distribution
$p_{\theta}^{\mathrm{C}}(\mathbf{x}_0\mid\mathbf{y},\mathbf{M})$ 
from masked training examples. Following the EDM formulation, a clean field is
perturbed according to
\begin{equation}
  \mathbf{x}_{\sigma}
  =
  \mathbf{x}_0+\sigma\boldsymbol{\epsilon},
  \qquad
  \boldsymbol{\epsilon}\sim\mathcal{N}(\mathbf{0},\mathbf{I}),
  \qquad
  \log\sigma\sim
  \mathcal{N}\!\left(P_{\mathrm{mean}},P_{\mathrm{std}}^2\right).
  \label{eq:edm_corruption}
\end{equation}
The preconditioned conditional denoiser 
$D_{\theta}^{\mathrm{C}}
(\mathbf{x}_{\sigma},\sigma\mid\mathbf{y},\mathbf{M})$
is trained using the noise-weighted objective
\begin{equation}
  \mathcal{L}_{\mathrm{C}}
  =
  \mathbb{E}_{\mathbf{x}_0,\mathbf{M},\sigma,\boldsymbol{\epsilon}}
  \left[
    \lambda(\sigma)
    \operatorname{MSE}\!\left(
      D_{\theta}^{\mathrm{C}}
      (\mathbf{x}_{\sigma},\sigma\mid\mathbf{y},\mathbf{M}),
      \mathbf{x}_0
    \right)
  \right].
  \label{eq:conditional_edm_loss}
\end{equation}
The weighting function $\lambda(\sigma)$ and denoiser preconditioning follow the EDM formulation.

At inference, let
$\sigma_0>\sigma_1>\cdots>\sigma_N=0$ denote the sampling noise levels.
Sampling begins from
$\mathbf{x}(\sigma_0)=\sigma_0\boldsymbol{\epsilon}$, where
$\boldsymbol{\epsilon}\sim\mathcal{N}(\mathbf{0},\mathbf{I})$, and follows
the conditional EDM probability-flow ODE
\begin{equation}
  \frac{\mathrm{d}\mathbf{x}(\sigma)}{\mathrm{d}\sigma}
  =
  \frac{
    \mathbf{x}(\sigma)
    -
    D_{\theta}^{\mathrm{C}}
    \!\left(
      \mathbf{x}(\sigma),\sigma
      \mid\mathbf{y},\mathbf{M}
    \right)
  }{\sigma}.
  \label{eq:conditional_edm_ode}
\end{equation}
The ODE is integrated from high noise to the clean-data limit using a
second-order Euler--Heun solver. The same observation pair
$(\mathbf{y},\mathbf{M})$ is supplied at every denoising evaluation.
Independent draws of $\boldsymbol{\epsilon}$ produce distinct conditional
reconstructions for the same observation pair.
The noise schedule and sampler configuration are reported in Appendix~\ref{sec:appendix_training}.

\subsection{Prior-guided EDM (G-EDM)}
\label{sec:guided_edm}

G-EDM separates full-field prior learning from observation conditioning. Its
denoiser $D_{\theta}^{\mathrm{P}}$ is trained on complete fields using the
same corruption process, loss weighting, and denoiser preconditioning as
C-EDM, but receives neither $\mathbf{y}$ nor $\mathbf{M}$. The training
objective is
\begin{equation}
  \mathcal{L}_{\mathrm{P}}
  =
  \mathbb{E}_{\mathbf{x}_0,\sigma,\boldsymbol{\epsilon}}
  \left[
    \lambda(\sigma)
    \operatorname{MSE}\!\left(
      D_{\theta}^{\mathrm{P}}(\mathbf{x}_{\sigma},\sigma),
      \mathbf{x}_0
    \right)
  \right].
  \label{eq:prior_edm_loss}
\end{equation}
The resulting model represents the full-field distribution
$p_{\theta}^{\mathrm{P}}(\mathbf{x}_0)$ rather than an
observation-conditioned distribution.

Prior sampling follows Equation~\ref{eq:conditional_edm_ode}, with
$D_{\theta}^{\mathrm{C}}$ replaced by $D_{\theta}^{\mathrm{P}}$. Starting
from the same Gaussian initialization as C-EDM, each sampling step first
advances the prior ODE using the Euler--Heun scheme and then applies an
observation-guidance correction. The guidance procedure is adapted from
DiffusionPDE~\cite{huang2024diffusionpde} and motivated by diffusion posterior
sampling~\cite{chung2023dps}.

At sampling step $i$, let
$\mathbf{x}^{(i)}=\mathbf{x}(\sigma_i)$ denote the current state, and let
$\widetilde{\mathbf{x}}^{(i+1)}$ denote the provisional state obtained by
advancing the prior ODE from $\sigma_i$ to $\sigma_{i+1}$. The denoised
estimate used for guidance is
\begin{equation*}
  \widehat{\mathbf{x}}_0^{(i)}
  =
  D_{\theta}^{\mathrm{P}}
  \!\left(\mathbf{x}^{(i)},\sigma_i\right).
\end{equation*}
Observation consistency is measured by
\begin{equation}
  \mathcal{L}_{\mathrm{obs}}^{(i)}
  =
  \sum_{c=1}^{C}
  \zeta_c
  \left\|
    \mathbf{M}\odot
    \left(
      \widehat{\mathbf{x}}_{0,c}^{(i)}-\mathbf{y}_c
    \right)
  \right\|_2,
  \label{eq:observation_guidance_loss}
\end{equation}
where $\zeta_c$ controls the relative contribution of channel $c$. The mask
restricts this unsquared $\ell_2$ discrepancy to observed locations.

Because $\mathcal{L}_{\mathrm{obs}}^{(i)}$ depends on
$\mathbf{x}^{(i)}$ through the prior denoiser, its gradient is backpropagated
through $D_{\theta}^{\mathrm{P}}$. The provisional state is corrected as
\begin{equation}
  \mathbf{x}^{(i+1)}
  =
  \widetilde{\mathbf{x}}^{(i+1)}
  -
  r_i
  \nabla_{\mathbf{x}^{(i)}}
  \mathcal{L}_{\mathrm{obs}}^{(i)},
  \label{eq:guidance_correction}
\end{equation}
where $r_i$ is the step-dependent guidance multiplier. The corrected state
$\mathbf{x}^{(i+1)}$ is then used as the input to the next Euler--Heun step.
The PDE residual used in DiffusionPDE is omitted, leaving
$\mathcal{L}_{\mathrm{obs}}^{(i)}$ as the sole guidance objective. The final
guidance configurations are summarized in
Appendix~\ref{sec:appendix_training}, and their selection is described in
Appendix~\ref{sec:appendix_guidance}. Independent initial noise fields
produce distinct observation-guided reconstructions for the same observation
pair $(\mathbf{y},\mathbf{M})$.

\FloatBarrier

\section{Experimental design and evaluation}
\label{sec:protocol}

\subsection{Benchmarks and observation protocol}
\label{sec:benchmarks}

We consider three benchmarks with four reconstruction targets. The Poisson
benchmark is posed on $\Omega=(0,1)^2$ as
\begin{equation}
  -\Delta u(\mathbf{s})=f(\mathbf{s}),
  \qquad
  \mathbf{s}\in\Omega,
  \qquad
  u|_{\partial\Omega}=0.
  \label{eq:poisson}
\end{equation}
We jointly reconstruct the solution $u$ and source $f$, with
$k_c\in\{1,2,4,6\}$ controlling the frequency content of the source.

The Navier--Stokes (NS) benchmark is governed by the two-dimensional
incompressible vorticity equation on the periodic unit square $(0,1)^2$,
\begin{equation}
  \frac{\partial\omega}{\partial t}
  +
  \mathbf{v}\cdot\nabla\omega
  =
  \nu\Delta\omega+q(\mathbf{s}),
  \qquad
  \mathbf{s}\in(0,1)^2,
  \qquad
  \nabla\cdot\mathbf{v}=0.
  \label{eq:navier_stokes}
\end{equation}
Here, $\mathbf{v}$ denotes velocity,
$\omega=\nabla\times\mathbf{v}$ is the reconstructed scalar vorticity, and
$q$ is a fixed forcing. We use ten-frame vorticity trajectories from the
Fourier neural operator dataset~\cite{li2021fno} at
$\nu\in\{10^{-3},10^{-4},10^{-5}\}$, corresponding to
$\mathrm{Re}=\nu^{-1}\in\{10^3,10^4,10^5\}$.

The Kuramoto--Sivashinsky (KS) benchmark is governed on the periodic domain
$x\in[0,L]$ by
\begin{equation}
  \frac{\partial u}{\partial t}
  +
  u\frac{\partial u}{\partial x}
  +
  \frac{\partial^2u}{\partial x^2}
  +
  \nu\frac{\partial^4u}{\partial x^4}
  =
  0.
  \label{eq:kuramoto_sivashinsky}
\end{equation}
Here, $u(t,x)$ denotes the evolving scalar field and $\nu$ is the viscosity
parameter. We reconstruct $u(t,x)$ from spatiotemporal windows in the
PDE-Refiner data splits~\cite{lippe2023pderefiner}, with
$\nu\in[0.5,1.5]$ grouped into five intervals. The domain length $L$ follows
the released trajectory metadata. Across the three benchmarks, higher $k_c$,
higher $\mathrm{Re}$, and lower $\nu$ correspond to greater field complexity.

U-Net and C-EDM are trained with uniformly sampled masks, whereas G-EDM is
trained on complete fields without observation masks. For each U-Net and
C-EDM training example, the observation fraction is drawn uniformly from
$\{5,8,10,15,20\}\%$, after which the observed locations are sampled
without replacement. All three methods are evaluated at the same five
observation fractions. Poisson uses the same observed locations for $u$ and
$f$, while NS and KS use time-invariant spatial sensor locations within each
temporal window. Fixed validation and test masks are shared across methods.

Additional observation-fraction tests include interpolation at 12\% and
extrapolation below the training range at 3\%. Poisson source-frequency
generalization uses $k_c=3.5$ for interpolation and $k_c=8$ for extrapolation,
both evaluated at the same five observation fractions. Mask-distribution
shifts are evaluated for Poisson and NS at a fixed 5\% observation count using
Gaussian--uniform, directional, and patch-missing masks. Each shifted mask is
paired with a uniform mask on the same field or trajectory. Dataset
construction, evaluation panels, and shifted mask distributions are detailed
in Appendices~\ref{sec:appendix_datasets},
\ref{sec:appendix_observation_panels}, and
\ref{sec:appendix_robustness}.

\subsection{Controlled comparison and ensemble statistics}
\label{sec:comparison_protocol}

Within each benchmark, all three methods use identical data splits, normalization statistics, and test observations.
The U-Net predictor and the
two EDM denoisers use matched backbone architectures. Sampling configurations for both EDM formulations are selected
using validation performance.
Architecture, training, and final sampling
configurations are reported in Appendix~\ref{sec:appendix_training}, while
the selection of G-EDM guidance parameters is detailed in
Appendix~\ref{sec:appendix_guidance}.

At test time, each EDM formulation generates
$N_{\mathrm{ens}}=32$ reconstructions,
$\{\widehat{\mathbf{x}}_0^{(m)}\}_{m=1}^{N_{\mathrm{ens}}}$. Their ensemble
mean is
\begin{equation}
  \widehat{\mathbf{x}}_0^{(\mathrm{E})}
  =
  \frac{1}{N_{\mathrm{ens}}}
  \sum_{m=1}^{N_{\mathrm{ens}}}
  \widehat{\mathbf{x}}_0^{(m)}.
  \label{eq:ensemble_mean}
\end{equation}
For a scalar reconstruction metric $d$, we report
\begin{equation}
  d_{\mathrm{E}}
  =
  d\!\left(\widehat{\mathbf{x}}_0^{(\mathrm{E})},\mathbf{x}_0\right),
  \qquad
  d_{\mathrm{S}}
  =
  \frac{1}{N_{\mathrm{ens}}}
  \sum_{m=1}^{N_{\mathrm{ens}}}
  d\!\left(\widehat{\mathbf{x}}_0^{(m)},\mathbf{x}_0\right).
  \label{eq:ensemble_statistics}
\end{equation}
The suffixes (E) and (S) denote ensemble-mean and mean memberwise statistics,
respectively. A ``sample'' refers to one generated member. The ensemble-size
analysis is provided in 
Appendix~\ref{sec:appendix_ensemble_size}.

\subsection{Evaluation metrics}
\label{sec:evaluation_criteria}

Before evaluating reconstruction and spectral accuracy, predicted values at
observed locations are replaced with the corresponding measurements.
Probabilistic metrics are computed only at unobserved locations. Reconstruction
accuracy is primarily evaluated using the full-field relative $\ell_2$ error
\begin{equation}
  \varepsilon_{\mathrm{rel}}
  =
  \frac{
    \left\|
      \widehat{\mathbf{x}}_0^{\mathrm{eval}}-\mathbf{x}_0
    \right\|_2
  }{
    \left\|\mathbf{x}_0\right\|_2
  },
  \label{eq:relative_l2}
\end{equation}
where $\widehat{\mathbf{x}}_0^{\mathrm{eval}}$ denotes the reconstruction
after restoring the observed values.

Frequency-domain evaluation separates agreement with the spectral
coefficients of the ground truth from agreement with its spectral power
distribution. Let
$\mathbf{X}=\mathcal{T}\mathbf{x}_0$ and
$\widehat{\mathbf{X}}
=\mathcal{T}\widehat{\mathbf{x}}_0^{\mathrm{eval}}$,
where $\mathcal{T}$ is the benchmark-specific orthonormal spatial transform.
For a set of spatial modes $\mathcal{A}$, define
\begin{equation*}
  \|\mathbf{Z}\|_{2,\mathcal{A}}^2
  =
  \sum_{t\in\mathcal{I}_t}
  \sum_{\kappa\in\mathcal{A}}
  |Z(t,\kappa)|^2,
\end{equation*}
where the sum over time is omitted for the static Poisson fields. The
phase-aligned coefficient error in frequency band $b$ is
\begin{equation}
  \varepsilon_b^{\mathrm{spec}}
  =
  \frac{
    \left\|\widehat{\mathbf{X}}-\mathbf{X}\right\|_{2,\mathcal{K}_b}
  }{
    \left\|\mathbf{X}\right\|_{2,\mathcal{K}}
  },
  \label{eq:bandwise_spectral_error}
\end{equation}
where $\mathcal{K}_b$ contains the spatial modes in band $b$ and
$\mathcal{K}$ contains all evaluated spatial modes. The common full-spectrum denominator preserves each band's
contribution to the total coefficient error.

The complementary metric evaluates normalized spectral shape. Given the power
spectrum $P(k)$, we define
$\widetilde{P}(k)=P(k)/\sum_j P(j)$ and
$\widetilde{P}_{\epsilon}(k)
=\max\{\widetilde{P}(k),\epsilon\}$, where
$\epsilon=10^{-14}$ prevents undefined logarithms. The bandwise log-power
error is
\begin{equation}
  D_b^{\log P}
  =
  \frac{1}{|\mathcal{B}_b|}
  \sum_{k\in\mathcal{B}_b}
  \left|
    \log_{10}
    \frac{
      \widetilde{P}_{\mathrm{pred},\epsilon}(k)
    }{
      \widetilde{P}_{\mathrm{true},\epsilon}(k)
    }
  \right|,
  \label{eq:bandwise_log_power_error}
\end{equation}
where $\mathcal{B}_b$ contains the wavenumbers in band $b$. Because $D_b^{\log P}$ depends on spectral power rather than spectral coefficients, it is insensitive to phase.

Predictive uncertainty is evaluated at unobserved locations using the
continuous ranked probability score (CRPS), empirical coverage of the central
90\% prediction interval, and mean interval width. CRPS measures the accuracy
of the ensemble predictive distribution, while coverage and width characterize
its calibration and sharpness. Details of the spectral transforms, frequency
bands, and probabilistic metrics are provided in Appendices~\ref{sec:appendix_metrics}
and~\ref{sec:appendix_spectral_results}.

\section{Results}
\label{sec:results}

\subsection{Reconstruction accuracy is field- and regime-dependent}
\label{sec:point_accuracy}

Table~\ref{tab:overall_full_relative_l2} compares the mean full-field relative $\ell_2$ errors across the four reconstruction targets, averaged over field-complexity settings and observation fractions. 
Consistent with the convexity of the relative $\ell_2$ metric, the ensemble-mean error was lower than the corresponding mean memberwise error for both EDM formulations.
C-EDM (E) was only marginally more accurate than U-Net for Poisson $u$, whereas G-EDM (E) led by a more pronounced numerical margin for NS $\omega$. U-Net performed best for Poisson $f$ and KS $u$. 
Overall, reconstruction accuracy was field-dependent, with no formulation consistently superior.

\begin{table}[!t]
  \centering
  \small
  \setlength{\tabcolsep}{4.2pt}
  \tbl{Mean full-field relative $\ell_2$ error (\%). Best and second-best results are shown in bold and underlined, respectively.}{%
  \begin{tabular}{lccccc}
    \toprule
    & \multicolumn{5}{c}{Full relative $\ell_2$ error (\%) $\downarrow$} \\
    \cmidrule(lr){2-6}
    Field & U-Net & C-EDM (E) & C-EDM (S) & G-EDM (E) & G-EDM (S) \\
    \midrule
    Poisson solution $u$ & \underline{0.2132} & \textbf{0.2078} & 0.2980 & 0.3705 & 0.4905 \\
    Poisson source $f$ & \textbf{8.3462} & \underline{8.9191} & 12.5897 & 11.2882 & 13.6826 \\
    NS vorticity $\omega$ & 1.9707 & 2.1200 & 2.8550 & \textbf{1.4318} & \underline{1.8343} \\
    KS state $u$ & \textbf{2.0630} & \underline{2.1991} & 3.2794 & 4.5670 & 6.6020 \\
    \bottomrule
  \end{tabular}}
  \label{tab:overall_full_relative_l2}
\end{table}

\FloatBarrier
\begin{figure}[!t]
  \centering
  \includegraphics[width=\textwidth]{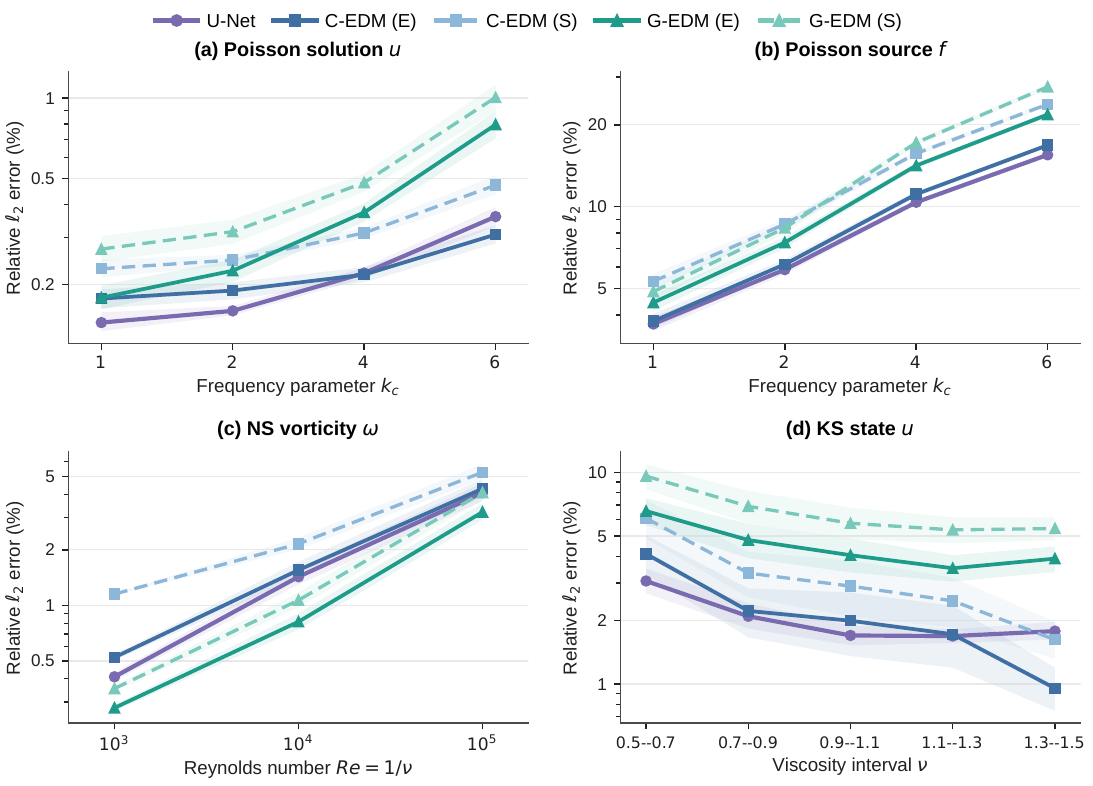}
  \caption{Mean full-field relative $\ell_2$ error across field-complexity regimes:
  (a) Poisson $u$, (b) Poisson $f$, (c) NS $\omega$, and
  (d) KS $u$. Values are averaged over the five observation
  fractions at each complexity setting; shading shows 95\% confidence intervals over independent fields
  or trajectories.}
  \label{fig:field_complexity}
\end{figure}

Figure~\ref{fig:field_complexity} resolves the aggregate results in Table~\ref{tab:overall_full_relative_l2} by field complexity.
Reconstruction error increased monotonically with $k_c$ for both Poisson fields and with $Re$ for NS; KS was generally more difficult at lower viscosity, although the trend was not strictly monotonic. 
For Poisson $u$, the leading method shifted from U-Net at $k_c=1$ and 2 to C-EDM (E) at $k_c=4$ and 6. 
KS showed the opposite pattern: U-Net led in the four lower-viscosity intervals, whereas C-EDM (E) led only at $\nu\in[1.3,1.5)$. 
U-Net remained best for Poisson $f$, and G-EDM (E) for NS $\omega$, across all tested complexity settings. 
Thus, increasing field complexity did not systematically increase the advantage of either EDM formulation over U-Net.

Reconstruction accuracy also depended on observation density (Appendix
Figure~\ref{fig:appendix_observation_fraction}). 
Across the five benchmark observation fractions, full-field errors generally decreased as more measurements were provided.
U-Net remained best for Poisson $f$, and G-EDM (E) for NS $\omega$, across all five fractions. 
For Poisson $u$, C-EDM (E) led at 5\% observations before U-Net took the lead from 8\% onward; KS showed the opposite transition, from U-Net at 5\% to C-EDM (E) from 8\% onward.
Together, the complexity- and density-dependent crossovers were field-specific: neither increasing complexity nor reducing observation density systematically increased the advantage of the EDM formulations over U-Net.

\begin{figure}[!t]
  \centering
  \includegraphics[width=\textwidth]{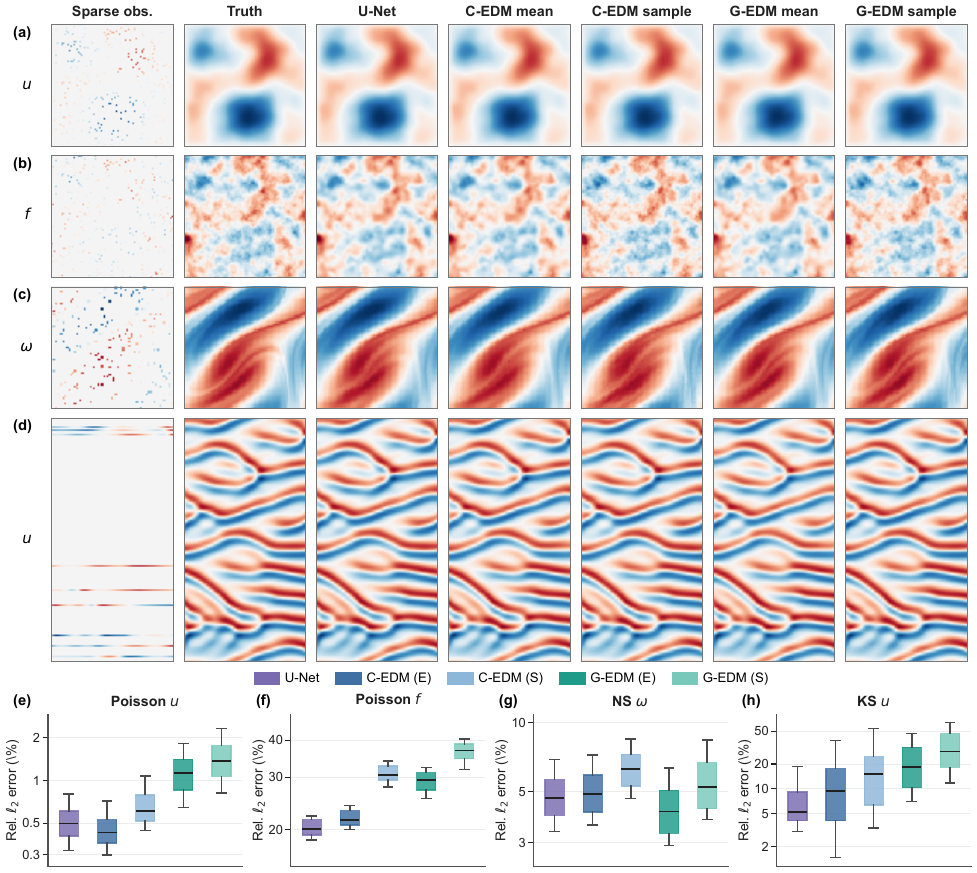}
  \caption{Reconstructions and error distributions for the most challenging setting in the main evaluation: 5\% observations with $k_c=6$ for Poisson, $Re=10^5$ for NS, and $\nu\in[0.5,0.7)$ for KS. Panels (a)--(d) compare sparse observations, ground truth, and reconstructions for one randomly selected test case per target; panels (e)--(h) summarize the paired full-field relative $\ell_2$ errors across all 50 test cases.}
  \label{fig:reconstruction_overview}
\end{figure}

Figure~\ref{fig:reconstruction_overview} examines the most challenging setting in the main evaluation for each benchmark: 5\% observations with $k_c=6$ for Poisson, $Re=10^5$ for NS, and $\nu\in[0.5,0.7)$ for KS.
The paired error distributions identify the same leading method for each reconstruction target as the aggregate results in Table~\ref{tab:overall_full_relative_l2}. Despite differences in full-field error, the representative reconstructions recover broadly similar dominant structures. This apparent large-scale agreement motivates the frequency-resolved analysis in Section~\ref{sec:ensemble_spectral_effects}, which examines whether the methods differ in their recovery of finer-scale content.

\FloatBarrier

\subsection{Ensemble averaging improves phase-aligned accuracy but can attenuate spectral tails}
\label{sec:ensemble_spectral_effects}

\begin{figure}[!t]
  \centering
  \includegraphics[width=\textwidth]{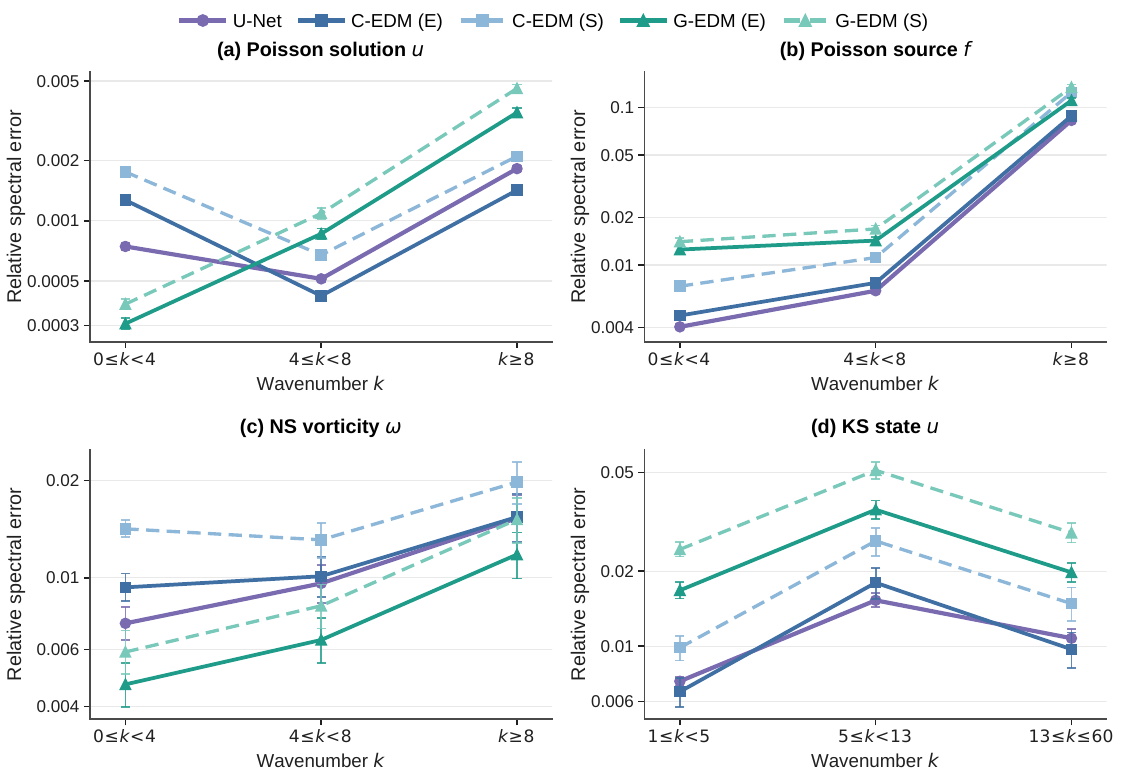}
  \caption{Mean bandwise phase-aligned spectral coefficient error $\varepsilon_b^{\mathrm{spec}}$ for the four reconstruction targets. Values are averaged over field-complexity settings and the five observation fractions in the main evaluation; error bars show paired-bootstrap 95\% confidence intervals.}
  \label{fig:bandwise_error}
\end{figure}

We determined dataset-specific low-, mid-, and high-frequency bands from the ground-truth spectral energy distributions before comparing the reconstruction methods (Appendix Figure~\ref{fig:truth_spectra}). Figure~\ref{fig:bandwise_error} reports the phase-aligned spectral coefficient error $\varepsilon_b^{\mathrm{spec}}$ defined in Equation~(16) over these bands. Consistent with the convexity of this metric, the ensemble-mean error was 10.9--37.6\% lower than the corresponding mean memberwise error across both EDM formulations, all four reconstruction targets, and all three frequency bands.

\begin{figure}[!t]
  \centering
  \includegraphics[width=\textwidth]{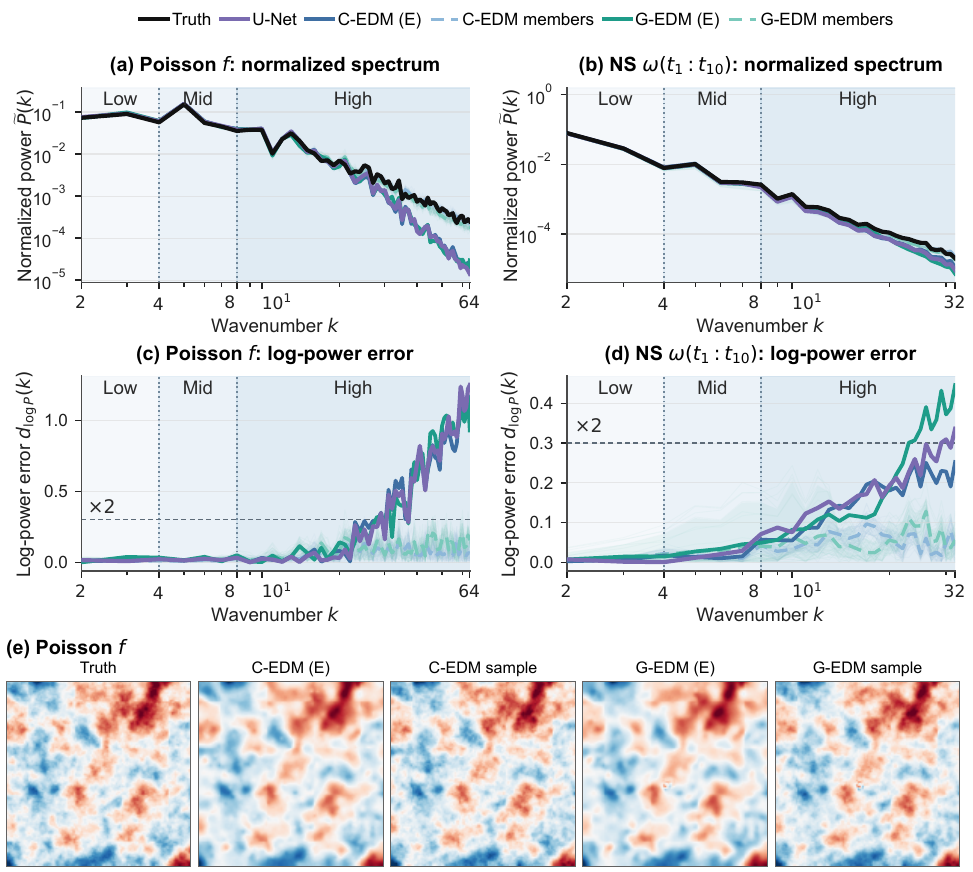}
  \caption{Spectral diagnostics and reconstructions for selected cases with 5\% observations: Poisson $f$ at $k_c=6$ and NS $\omega$ at $Re=10^5$. Panels (a,b) show normalized spectra, and panels (c,d) show per-wavenumber log-power errors. Faint curves represent individual EDM members. Highlighted curves show their pointwise median spectra in (a,b) and mean errors in (c,d), with 10th--90th percentile error bands. Panel (e) compares the Poisson $f$ ground truth with the ensemble mean and one member from each EDM formulation.}
  \label{fig:representative_spectra}
\end{figure}

Table~\ref{tab:ensemble_spectrum_effect} reports the high-frequency log-power error $D_H^{\log P}$ defined in Equation~(17) for Poisson $f$ and NS $\omega$. For Poisson $f$, both EDM memberwise errors were lower than the U-Net error at every $k_c$. The C-EDM memberwise error was about one fifth of its ensemble-mean counterpart, while the memberwise advantage within G-EDM widened as $k_c$ increased. For NS, the reversal occurred only at $Re=10^5$, where both EDM memberwise errors were also lower than their ensemble-mean counterparts and the U-Net error. No high-frequency reversal occurred for Poisson $u$ or KS. Complete bandwise results are reported in Appendix Tables~\ref{tab:appendix_poisson_spectrum}--\ref{tab:appendix_ks_spectrum}. Thus, in selected regimes, individual EDM members on average matched high-wavenumber power more closely than both their ensemble means and U-Net, despite the lower phase-aligned error achieved by ensemble averaging.

\begin{table}[!ht]
  \centering
  \small
  \setlength{\tabcolsep}{4.2pt}
  \tbl{High-frequency ($k\geq8$) log-power error $D_{\mathrm{H}}^{\log P}$ across $k_c$ for Poisson $f$ and $Re$ for NS $\omega$, averaged over test cases and observation fractions. Best and second-best results are shown in bold and underlined, respectively.}{%
  \begin{tabular}{llccccc}
    \toprule
    & & \multicolumn{5}{c}{Mean absolute log-power error $D_{\mathrm{H}}^{\log P}$ $\downarrow$} \\
    \cmidrule(lr){3-7}
    Field & Regime & U-Net & C-EDM (E) & C-EDM (S) & G-EDM (E) & G-EDM (S) \\
    \midrule
    Poisson $f$ & $k_c=1$ & $0.3586$ & $0.3348$ & $\mathbf{0.0630}$ & $0.2235$ & $\underline{0.1509}$ \\
     & $k_c=2$ & $0.3356$ & $0.3170$ & $\mathbf{0.0600}$ & $0.2460$ & $\underline{0.1447}$ \\
     & $k_c=4$ & $0.2940$ & $0.2918$ & $\mathbf{0.0570}$ & $0.2392$ & $\underline{0.0926}$ \\
     & $k_c=6$ & $0.2703$ & $0.2778$ & $\mathbf{0.0557}$ & $0.2256$ & $\underline{0.0704}$ \\
    \addlinespace[2pt]
    NS $\omega$ & $Re=10^{3}$ & $\underline{6.2367}$ & $\mathbf{6.1541}$ & $6.8390$ & $6.2545$ & $6.4540$ \\
     & $Re=10^{4}$ & $\underline{0.5889}$ & $0.6137$ & $0.7922$ & $\mathbf{0.5128}$ & $0.6448$ \\
     & $Re=10^{5}$ & $0.1674$ & $0.1646$ & $\underline{0.0625}$ & $0.1204$ & $\mathbf{0.0371}$ \\
    \bottomrule
  \end{tabular}}
  \label{tab:ensemble_spectrum_effect}
\end{table}

Figure~\ref{fig:representative_spectra} examines this effect in randomly selected cases at 5\% observations for Poisson $f$ at $k_c=6$ and NS $\omega$ at $Re=10^5$. The faster high-wavenumber decay of the ensemble-mean spectrum is clear for Poisson $f$ in panel (a). In both cases, the EDM members have lower spectral-tail log-power errors than their ensemble-mean counterparts in panels (c) and (d). As shown in panel (e), the Poisson $f$ ensemble means are also smoother than the individual EDM members, which retain more fine-scale texture. 

The two spectral metrics in Figure~\ref{fig:bandwise_error} and Table~\ref{tab:ensemble_spectrum_effect} capture different aspects of reconstruction. The phase-aligned coefficient error compares Fourier coefficients with the ground truth, whereas the log-power error compares normalized spectral power without regard to phase. A lower log-power error therefore indicates closer agreement in high-wavenumber power but not necessarily the correct spatial placement of the corresponding fine-scale structures. The relative $\ell_2$ distance between normalized power spectra produced a different ordering from the high-frequency log-power comparison (Appendix Table~\ref{tab:appendix_normalized_power_spectrum}). This distance emphasizes energetic modes and is less sensitive to the low-power spectral tail. Overall, ensemble averaging consistently reduced phase-aligned error but attenuated high-wavenumber power in selected fields and regimes, possibly because member-specific components cancel during averaging.

\FloatBarrier

\subsection{Prior-guided generation is least sensitive to observation-mask distribution shifts}
\label{sec:robustness}

Method rankings remained unchanged under the observation-fraction and Poisson source-frequency shifts shown in Appendix Figures~\ref{fig:appendix_observation_fraction} and \ref{fig:appendix_poisson_frequency_generalization}.
Clearer differences emerged under shifts in the spatial mask distribution. At 5\% observations, we compared uniform masks with Gaussian--uniform, directional, and patch-missing masks on paired Poisson fields and NS trajectories while holding all model and sampler settings fixed (Figure~\ref{fig:mask_examples}). Their construction is detailed in Appendix~\ref{sec:appendix_robustness}. KS was excluded because its fixed one-dimensional sensor layout has no direct analogue of these two-dimensional mask shifts.

\begin{figure}[!t]
  \centering
  \includegraphics[width=\textwidth]{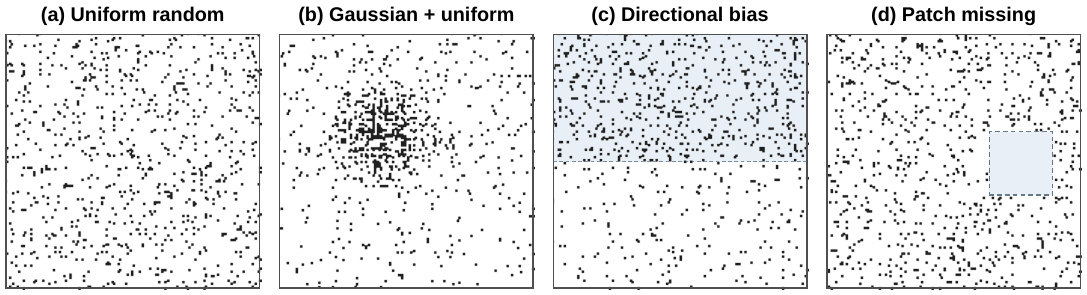}
  \caption{Mask distributions at a fixed 5\% observation count. Gaussian--uniform sampling combines clustered and uniformly sampled sensors, directional sampling concentrates sensors in one half of the domain, and patch missingness leaves a square region unobserved.}
  \label{fig:mask_examples}
\end{figure}

\begin{figure}[!t]
  \centering
  \includegraphics[width=\textwidth]{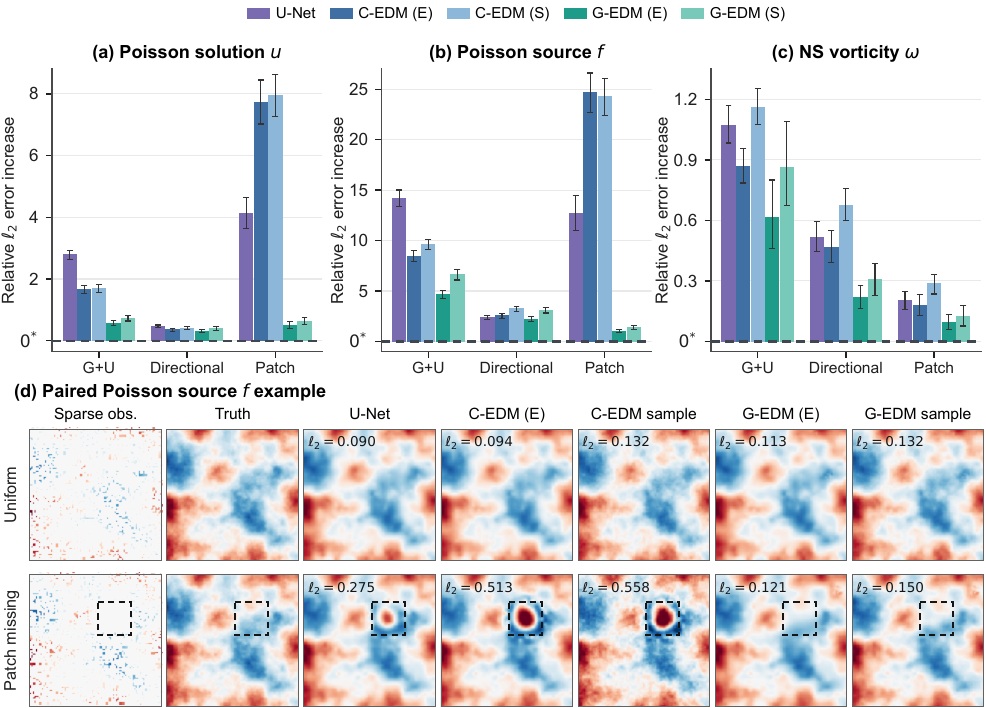}
  \caption{Sensitivity to mask-distribution shifts at 5\% observations. Panels (a)--(c) show the mean paired increase (percentage points) in full-field relative $\ell_2$ error from each method's uniform-mask baseline, with 95\% bootstrap confidence intervals. Panel (d) compares reconstructions of the same Poisson $f$ test field under uniform and patch-missing masks. Annotations give full-field relative $\ell_2$ errors.}
  \label{fig:mask_robustness}
\end{figure}

Figure~\ref{fig:mask_robustness}(a)--(c) quantifies sensitivity by the mean paired increase in full-field relative $\ell_2$ error from each method's uniform-mask baseline, $\Delta\varepsilon_{m,g}=\varepsilon_{m,g}-\varepsilon_{m,\mathrm{uniform}}$. G-EDM (E) had the smallest increase across all nine combinations of reconstructed fields and shifted masks. The largest separation occurred for Poisson $f$ under patch missingness, where the mean error increased by 12.72 percentage points for U-Net, 24.68 for C-EDM (E), and only 1.01 for G-EDM (E). The differences were smaller for the directional Poisson masks and across all three NS masks, but G-EDM (E) remained the least affected.
Figure~\ref{fig:mask_robustness}(d) illustrates this result using paired Poisson $f$ reconstructions under uniform and patch-missing masks. For this case, the full-field relative $\ell_2$ error increased from 0.090 to 0.275 for U-Net and from 0.094 to 0.513 for C-EDM (E), whereas it changed only from 0.113 to 0.121 for G-EDM (E). The reconstructions indicate that much of the additional U-Net and C-EDM (E) error arose within the unobserved patch. Additional paired examples under Gaussian--uniform and directional masks are provided in Appendix Figure~\ref{fig:appendix_mask_cases}.

The absolute-error rankings across the shifted test sets broadly followed the sensitivity results. G-EDM (E) achieved the lowest mean absolute error in seven of the nine shifted conditions (Appendix Table~\ref{tab:appendix_mask_robustness}). The directional Poisson masks were the two exceptions, favoring C-EDM (E) for $u$ and U-Net for $f$. Although G-EDM (E) was not the most accurate method for either Poisson variable under uniform masks, it became the most accurate for both under Gaussian--uniform and patch-missing masks.

The formulations differ in how they incorporate the observation mask. 
U-Net and C-EDM learn an amortized conditional mapping from observations and masks drawn from the uniform training distribution. 
G-EDM instead learns a full-field prior independently of the observations and applies the test mask through guidance during sampling. 
This separation between prior learning and observation enforcement helps explain the smaller degradation of G-EDM (E) when the spatial distribution of observations changed.

\subsection{Conditional generation generally provides more reliable uncertainty estimates}

Table~\ref{tab:uncertainty_quality} evaluates uncertainty at unobserved locations under paired 5\% uniform masks using CRPS, empirical 90\% interval coverage, and interval width. Lower CRPS is preferable, whereas a narrower interval is favorable only when coverage is comparable.
C-EDM achieved lower CRPS for Poisson $u$, Poisson $f$, and KS, and its coverage was closer to the nominal 90\% level for all four reconstruction targets. 
For Poisson $f$, G-EDM produced a narrower interval, but its coverage was only 60.2\%, compared with 85.9\% for C-EDM.
C-EDM led on all three diagnostics for KS, with lower CRPS, a narrower interval, and coverage closer to 90\%. 
NS was the exception, with G-EDM achieving lower CRPS and a narrower interval at similar coverage.

\begin{table}[!t]
  \centering
  \small
  \setlength{\tabcolsep}{4.2pt}
  \tbl{Empirical uncertainty estimation at unobserved locations under paired 5\% uniform masks, computed from ensembles of $N_{\mathrm{ens}}=32$ members.}{%
  \begin{tabular}{llrccc}
    \toprule
    Field & Method & $N_{\mathrm{case}}$ & CRPS $\downarrow$ & Coverage (\%) $\to 90$ & 90\% width \\
    \midrule
    Poisson $u$ & C-EDM & 200 & \textbf{0.0013} & \textbf{84.9} & 0.0064 \\
     & G-EDM & 200 & 0.0020 & 61.2 & 0.0068 \\
    \addlinespace
    Poisson $f$ & C-EDM & 200 & \textbf{0.0656} & \textbf{85.9} & 0.3506 \\
     & G-EDM & 200 & 0.0871 & 60.2 & 0.3000 \\
    \addlinespace
    NS $\omega$ & C-EDM & 150 & 0.0158 & \textbf{87.6} & 0.0726 \\
     & G-EDM & 150 & \textbf{0.0105} & 87.2 & 0.0486 \\
    \addlinespace
    KS $u$ & C-EDM & 250 & \textbf{0.0220} & \textbf{93.9} & 0.1578 \\
     & G-EDM & 250 & 0.0411 & 79.1 & 0.3307 \\
    \bottomrule
  \end{tabular}}
  \label{tab:uncertainty_quality}
\end{table}

\begin{figure}[!t]
  \centering
  \includegraphics[width=\textwidth]{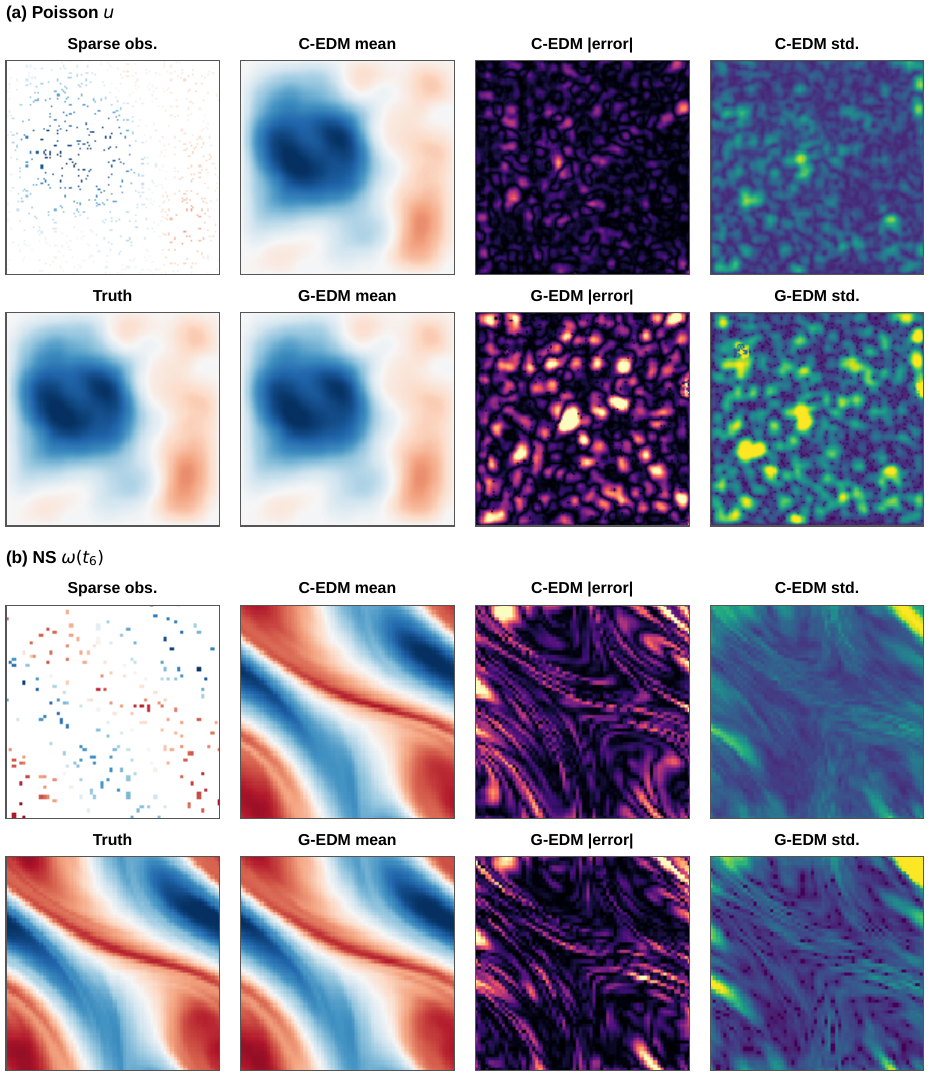}
  \caption{Spatial uncertainty for representative Poisson $u$ and NS $\omega(t_6)$ fields with 5\% uniformly sampled observations. Panels show observations, ground truth, ensemble means, absolute errors, and standard deviations for both EDM formulations ($N_{\mathrm{ens}}=32$). Errors and standard deviations are normalized by the RMS amplitude of the corresponding ground-truth field.}
  \label{fig:uncertainty_maps}
\end{figure}

Ensemble standard deviation provides a spatially resolved measure of predictive spread. Figure~\ref{fig:uncertainty_maps} compares its spatial distribution with absolute reconstruction error for representative Poisson $u$ and NS cases.
For both EDM formulations, regions of larger predictive spread broadly coincided with regions of larger reconstruction error.
Additional Poisson $f$ and KS examples in Appendix Figures~\ref{fig:appendix_uncertainty_poisson_f} and \ref{fig:appendix_uncertainty_ks} show similar spatial correspondence.
Together, Table~\ref{tab:uncertainty_quality} and Figure~\ref{fig:uncertainty_maps} characterize uncertainty through predictive diagnostics at individual locations and the spatial relationship between ensemble spread and reconstruction error.

\subsection{Prior-guided sampling is costly and sensitive to guidance parameters}

The number of G-EDM sampling steps and the guidance weights were selected separately for each benchmark on validation data, as detailed in Appendix~\ref{sec:appendix_guidance}. Validation error was sensitive to their combination, and neither stronger guidance nor more sampling steps consistently improved reconstruction accuracy. 
The selected configurations were $(N,\zeta_u,\zeta_f)=(400,10,20)$ for Poisson, $(N,\zeta)=(1{,}600,5)$ for NS, and $(N,\zeta)=(1{,}600,70)$ for KS. 
Increasing the number of steps for Poisson from 400 to 800 yielded no further improvement. 
For NS, increasing the number of steps from 1,600 to 2,000 reduced the validation error by only 1.3\% at 25\% greater sampling cost.

Under the final sampling configurations, Table~\ref{tab:sampling_cost} compares the per-member inference costs of the three reconstruction formulations. 
U-Net required one network evaluation for each reconstruction. 
C-EDM used 40 sampling steps for Poisson and NS and 20 steps for KS, corresponding to 79 and 39 denoiser evaluations per member. 
G-EDM used 400 sampling steps for Poisson and 1,600 steps for NS and KS, corresponding to 799--3,199 denoiser evaluations per member.
In addition, every G-EDM sampling step required one observation-gradient evaluation. The resulting per-member latency was 43--87 times the U-Net baseline for C-EDM and 1,439--6,604 times for G-EDM.
Compared with C-EDM, the per-member latency of G-EDM was approximately 20 times higher for Poisson, 75 times higher for NS, and 154 times higher for KS.

All ensemble-based results used $N_{\mathrm{ens}}=32$ for both EDM formulations. This common ensemble size was selected using the C-EDM validation results reported in Appendix~\ref{sec:appendix_ensemble_size}. Ensemble-mean reconstruction error showed diminishing improvement beyond 16 members, whereas some uncertainty estimates continued to improve, supporting the use of 32 members for the reported comparisons. The sampling trajectories are independent and can be evaluated in parallel along the batch dimension, although the denoising steps within each trajectory remain sequential. 
Consequently, the wall time of a 32-member ensemble is not simply 32 times the batch-size-one latency in Table~\ref{tab:appendix_ensemble_size}, but depends on the available batch parallelism and device memory.
Overall, C-EDM directly modeled the observation-conditioned distribution and produced ensemble reconstructions with far fewer sampling steps, while retaining competitive reconstruction accuracy and generally more reliable uncertainty estimates. G-EDM instead applied benchmark-specific observation guidance over longer sampling trajectories and showed lower sensitivity to observation-mask distribution shifts.

\begin{table}[!t]
  \centering
  \small
  \setlength{\tabcolsep}{3.8pt}
  \tbl{Per-member inference cost under the final sampling configurations, measured with a batch size of 1 on an NVIDIA RTX 4090. Relative latency is normalized to the corresponding U-Net latency.}{%
  \begin{tabular}{llrrrrr}
    \toprule
    Dataset & Method & Steps & Denoiser evals
      & Obs.-gradient evals
      & Latency (s) & Relative latency \\
    \midrule
    Poisson & U-Net & -- & 1   & 0     & 0.0140 & $1\times$ \\
            & C-EDM & 40 & 79  & 0     & 1.009 & $72\times$ \\
            & G-EDM & 400 & 799 & 400  & 20.17 & $1{,}439\times$ \\
    \addlinespace
    NS      & U-Net & -- & 1     & 0     & 0.0107 & $1\times$ \\
            & C-EDM & 40 & 79    & 0     & 0.936  & $87\times$ \\
            & G-EDM & 1,600 & 3,199 & 1,600 & 70.13 & $6{,}555\times$ \\
    \addlinespace
    KS      & U-Net & -- & 1     & 0     & 0.0140 & $1\times$ \\
            & C-EDM & 20 & 39    & 0     & 0.601  & $43\times$ \\
            & G-EDM & 1,600 & 3,199 & 1,600 & 92.26 & $6{,}604\times$ \\
    \bottomrule
  \end{tabular}}
  \label{tab:sampling_cost}
\end{table}

\section{Discussion}
\label{sec:discussion}

Reconstruction accuracy did not follow a universal ranking among deterministic regression, conditional generation, and prior-guided generation. 
The leading formulation varied with the reconstructed field and operating regime, and neither increasing field complexity nor reducing observation density consistently favored generative reconstruction. 
Problem difficulty alone therefore does not determine whether a generative method is preferable. 
Deterministic regression remains a strong choice when the objective is a single accurate reconstruction at low inference cost.
Diffusion models, however, can generate multiple plausible reconstructions from the same sparse observations, providing ensemble predictions and a natural basis for characterizing predictive uncertainty.

The ensemble mean and individual members represent different scientific outputs. 
Within both EDM formulations, ensemble averaging consistently reduced phase-aligned error, whereas individual members more closely matched high-wavenumber power in selected fields and regimes. 
This contrast may result from cancellation of member-specific high-wavenumber components during averaging. Since the log-power metric is phase-insensitive, the memberwise advantage concerns spectral amplitudes rather than the spatial placement of fine-scale structures. 
The ensemble mean is therefore more appropriate when pointwise agreement is primary, while the member distribution is more informative for predictive variability and phase-insensitive spectral statistics.

The two EDM formulations incorporate observations in fundamentally different ways. 
In conditional generation, C-EDM amortizes observation conditioning during training, enabling much shorter sampling trajectories.
Its uncertainty estimates were also generally more reliable.
In prior-guided generation, G-EDM learns a full-field prior independently of the observations and applies the observation mask through test-time guidance.
This separation is consistent with its lower sensitivity to the tested mask-distribution shifts. 
G-EDM nevertheless required longer sampling trajectories, substantially greater inference cost, and benchmark-specific guidance selection. 
The preferred formulation therefore depends on the required output, the anticipated observation process, and the available computational budget.

Several questions remain regarding more realistic observational and physical settings.
The mask-shift analysis covers limited spatial distributions, while the spatiotemporal masks remain fixed over time and the measurements are noise-free. Generalization to time-varying sensing, noisy observations, and three-dimensional fields therefore remains unresolved.
The evaluation also does not directly assess whether reconstructed fields satisfy governing equations, conservation laws, or boundary conditions. Future work should extend the comparison to noisy three-dimensional systems with broader spatial and temporal sensing geometries and enforce physical consistency.

\section{Conclusions}
\label{sec:conclusion}

In this work, we investigated when diffusion-based generative reconstruction is preferable to deterministic regression for physical-field reconstruction from sparse observations. To this end, we conducted a controlled comparison of three architecture-matched formulations: a deterministic U-Net, conditional EDM (C-EDM), and prior-guided EDM (G-EDM). The comparison covered four reconstruction targets from the 2D Poisson equation, 2D Navier--Stokes flow, and 1D Kuramoto--Sivashinsky dynamics, with controlled variations in field complexity, observation density, and mask distribution.

The results show that sparse physical-field reconstruction does not admit a universally superior deterministic or generative formulation. Reconstruction accuracy depended on the reconstructed field and operating regime, and neither greater field complexity nor sparser observations consistently favored diffusion models. Deterministic regression remained a strong low-cost option when the objective was a single accurate reconstruction. Diffusion models, however, provided ensembles of plausible reconstructions, making them useful when uncertainty, sample diversity, or robustness to changing sensing conditions was important.

The ensemble results further showed that different summaries of the generated distribution serve different purposes. Ensemble means achieved lower phase-aligned error and can be viewed as variance-reduced point estimates, whereas individual samples more closely preserved variability and high-wavenumber power in selected fields and regimes. The comparison between C-EDM and G-EDM also revealed distinct trade-offs in how observations are incorporated. C-EDM generally provided more reliable uncertainty estimates at substantially lower inference cost, while G-EDM was least sensitive to the tested mask-distribution shifts but required longer sampling and guidance-weight tuning.

Taken together, these findings clarify that the value of generative reconstruction lies not in universally better pointwise accuracy, but in how its distributional outputs are used. They provide practical guidance for selecting reconstruction methods according to the required scientific output, anticipated sensing conditions, and computational budget. They also suggest concrete directions for future sparse-reconstruction methods, including improving uncertainty estimation in deterministic approaches, preserving fine-scale sample variability in conditional generative models, and reducing the sampling cost and guidance sensitivity of prior-guided methods.

\FloatBarrier

\section*{Acknowledgments}

The work is supported by the National Natural Science Foundation of China (No. 62506367 and No. 62276269) and the Beijing Natural Science Foundation (No. F261002). 

\section*{Declaration of AI use}

During the preparation of this manuscript, H.Z. used OpenAI Codex (GPT-5.6, OpenAI) to improve the language, clarity, and flow of author-prepared text. All AI-assisted edits were critically reviewed and revised by the authors. The authors take full responsibility for the originality, accuracy, and integrity of the manuscript, including its references.

\section*{Data availability statement}

The data supporting the findings of this study are available from the
corresponding author upon reasonable request.

\section*{Code availability statement}

The source code used for model training, reconstruction, and evaluation is
available from the corresponding author upon reasonable request.

\section*{Disclosure statement}

No potential conflict of interest was reported by the authors.

\bibliographystyle{tfnlm}
\bibliography{references}

\begin{thebibliography}{10}
\providecommand{\url}[1]{\normalfont{#1}}
\providecommand{\urlprefix}{Available from: }

\bibitem{brunton2020machine}
Brunton~SL, Noack~BR, Koumoutsakos~P. Machine learning for fluid mechanics.
  Annual Review of Fluid Mechanics. 2020;\hspace{0pt}52:477--508.

\bibitem{lu2022partial}
Lu~PY, Ari{\~n}o~Bernad~J, Solja{\v{c}}i{\'c}~M. Discovering sparse
  interpretable dynamics from partial observations. Communications Physics.
  2022;\hspace{0pt}5:206.

\bibitem{course2023state}
Course~K, Nair~PB. State estimation of a physical system with unknown governing
  equations. Nature. 2023;\hspace{0pt}622:261--267.

\bibitem{berkooz1993pod}
Berkooz~G, Holmes~P, Lumley~JL. The proper orthogonal decomposition in the
  analysis of turbulent flows. Annual Review of Fluid Mechanics.
  1993;\hspace{0pt}25:539--575.

\bibitem{everson1995gappy}
Everson~R, Sirovich~L. Karhunen--lo{\`e}ve procedure for gappy data. Journal of
  the Optical Society of America A. 1995;\hspace{0pt}12(8):1657--1664.

\bibitem{willcox2006gappy}
Willcox~K. Unsteady flow sensing and estimation via the gappy proper orthogonal
  decomposition. Computers \& Fluids. 2006;\hspace{0pt}35(2):208--226.

\bibitem{donoho2006compressed}
Donoho~DL. Compressed sensing. IEEE Transactions on Information Theory.
  2006;\hspace{0pt}52(4):1289--1306.

\bibitem{callaham2019sparse}
Callaham~JL, Maeda~K, Brunton~SL. Robust flow reconstruction from limited
  measurements via sparse representation. Physical Review Fluids.
  2019;\hspace{0pt}4(10):103907.

\bibitem{erichson2020shallow}
Erichson~NB, Mathelin~L, Yao~Z, et~al. Shallow neural networks for fluid flow
  reconstruction with limited sensors. Proceedings of the Royal Society A:
  Mathematical, Physical and Engineering Sciences.
  2020;\hspace{0pt}476(2238):20200097.

\bibitem{nair2020leveraging}
Nair~NJ, Goza~A. Leveraging reduced-order models for state estimation using
  deep learning. Journal of Fluid Mechanics. 2020;\hspace{0pt}897:R1.

\bibitem{dubois2022flow}
Dubois~P, Gomez~T, Planckaert~L, et~al. Machine learning for fluid flow
  reconstruction from limited measurements. Journal of Computational Physics.
  2022;\hspace{0pt}448:110733.

\bibitem{fukami2019superres}
Fukami~K, Fukagata~K, Taira~K. Super-resolution reconstruction of turbulent
  flows with machine learning. Journal of Fluid Mechanics.
  2019;\hspace{0pt}870:106--120.

\bibitem{mo2025reconstructing}
Mo~Y, Magri~L. Reconstructing unsteady flows from sparse, noisy measurements
  with a physics-constrained convolutional neural network. Physical Review
  Fluids. 2025;\hspace{0pt}10(3):034901.

\bibitem{zhang2025energy}
Zhang~Q, Krotov~D, Karniadakis~GE. Operator learning for reconstructing flow
  fields from sparse measurements: An energy transformer approach. Journal of
  Computational Physics. 2025;\hspace{0pt}538:114148.

\bibitem{fukami2021voronoi}
Fukami~K, Maulik~R, Ramachandra~N, et~al. Global field reconstruction from
  sparse sensors with {Voronoi} tessellation-assisted deep learning. Nature
  Machine Intelligence. 2021;\hspace{0pt}3(11):945--951.

\bibitem{santos2023senseiver}
Santos~JE, Fox~ZR, Mohan~A, et~al. Development of the {Senseiver} for efficient
  field reconstruction from sparse observations. Nature Machine Intelligence.
  2023;\hspace{0pt}5(11):1317--1325.

\bibitem{luo2024continuous}
Luo~X, Xu~W, Nadiga~BT, et~al. Continuous field reconstruction from sparse
  observations with implicit neural networks. In: International Conference on
  Learning Representations; 2024.

\bibitem{guoze2026geometryaware}
Sun~G, Miao~T, Huang~H, et~al. Geometry-aware neural optimizer for shape
  optimization and inversion. In: Forty-third International Conference on
  Machine Learning; 2026.
  \urlprefix\url{https://openreview.net/forum?id=PTaUjEBHat}.

\bibitem{raissi2019physics}
Raissi~M, Perdikaris~P, Karniadakis~GE. Physics-informed neural networks: A
  deep learning framework for solving forward and inverse problems involving
  nonlinear partial differential equations. Journal of Computational Physics.
  2019;\hspace{0pt}378:686--707.

\bibitem{raissi2020hidden}
Raissi~M, Yazdani~A, Karniadakis~GE. Hidden fluid mechanics: Learning velocity
  and pressure fields from flow visualizations. Science.
  2020;\hspace{0pt}367(6481):1026--1030.

\bibitem{stuart2010inverse}
Stuart~AM. Inverse problems: A bayesian perspective. Acta Numerica.
  2010;\hspace{0pt}19:451--559.

\bibitem{kingma2014vae}
Kingma~DP, Welling~M. Auto-encoding variational bayes. In: International
  Conference on Learning Representations; 2014.
  \urlprefix\url{https://arxiv.org/abs/1312.6114}.

\bibitem{goodfellow2014gan}
Goodfellow~IJ, Pouget-Abadie~J, Mirza~M, et~al. Generative adversarial nets.
  In: Advances in Neural Information Processing Systems; Vol.~27; 2014.

\bibitem{ho2020ddpm}
Ho~J, Jain~AN, Abbeel~P. Denoising diffusion probabilistic models. In: Advances
  in Neural Information Processing Systems; Vol.~33; 2020. p. 6840--6851.

\bibitem{song2021score}
Song~Y, Sohl-Dickstein~J, Kingma~DP, et~al. Score-based generative modeling
  through stochastic differential equations. In: International Conference on
  Learning Representations; 2021.

\bibitem{lipman2023flowmatching}
Lipman~Y, Chen~RTQ, Ben-Hamu~H, et~al. Flow matching for generative modeling.
  In: International Conference on Learning Representations; 2023.

\bibitem{gundersen2021semiconditional}
Gundersen~K, Oleynik~A, Blaser~N, et~al. Semi-conditional variational
  auto-encoder for flow reconstruction and uncertainty quantification from
  limited observations. Physics of Fluids. 2021;\hspace{0pt}33(1):017119.

\bibitem{guemes2022raseedgan}
G{\"u}emes~A, Sanmiguel~Vila~C, Discetti~S. Super-resolution generative
  adversarial networks of randomly-seeded fields. Nature Machine Intelligence.
  2022;\hspace{0pt}4(12):1165--1173.

\bibitem{du2024confield}
Du~P, Parikh~MH, Fan~X, et~al. Conditional neural field latent diffusion model
  for generating spatiotemporal turbulence. Nature Communications.
  2024;\hspace{0pt}15:10416.

\bibitem{zhou2026perflow}
Zhou~H, Zhang~R, Wan~H, et~al. {PerFlow}: Physics-embedded rectified flow for
  efficient reconstruction and uncertainty quantification of spatiotemporal
  dynamics. arXiv.
  2026;\hspace{0pt}\urlprefix\url{https://arxiv.org/abs/2605.03548}.

\bibitem{oommen2026turbulence}
Oommen~V, Khodakarami~S, Bora~A, et~al. Learning turbulent flows with
  generative models for super resolution and sparse flow reconstruction. Nature
  Communications. 2026;\hspace{0pt}17:3707.

\bibitem{shysheya2024conditional}
Shysheya~A, Diaconu~C, Bergamin~F, et~al. On conditional diffusion models for
  {PDE} simulations. In: Advances in Neural Information Processing Systems;
  Vol.~37; 2024. p. 23246--23300.

\bibitem{baldassari2023conditional}
Baldassari~L, Siahkoohi~A, Garnier~J, et~al. Conditional score-based diffusion
  models for bayesian inference in infinite dimensions. In: Advances in Neural
  Information Processing Systems; Vol.~36; 2023. p. 24262--24290.

\bibitem{li2025palsb}
Li~Z, Dou~H, Fang~S, et~al. Physics-aligned field reconstruction with diffusion
  bridge. In: International Conference on Learning Representations; 2025.

\bibitem{chung2023dps}
Chung~H, Kim~J, McCann~MT, et~al. Diffusion posterior sampling for general
  noisy inverse problems. In: International Conference on Learning
  Representations; 2023.

\bibitem{shu2023physicsdiffusion}
Shu~D, Li~Z, Barati~Farimani~A. A physics-informed diffusion model for
  high-fidelity flow field reconstruction. Journal of Computational Physics.
  2023;\hspace{0pt}478:111972.

\bibitem{li2024s3gm}
Li~Z, Han~W, Zhang~Y, et~al. Learning spatiotemporal dynamics with a pretrained
  generative model. Nature Machine Intelligence.
  2024;\hspace{0pt}6(12):1566--1579.

\bibitem{huang2024diffusionpde}
Huang~J, Yang~G, Wang~Z, et~al. {DiffusionPDE}: Generative {PDE}-solving under
  partial observation. In: Advances in Neural Information Processing Systems;
  Vol.~37; 2024. p. 130291--130323.

\bibitem{amoros2026guiding}
Amor{\'o}s-Trepat~M, Medrano-Navarro~L, Liu~Q, et~al. Guiding diffusion models
  to reconstruct flow fields from sparse data. Physics of Fluids.
  2026;\hspace{0pt}38(1):015112.

\bibitem{ronneberger2015unet}
Ronneberger~O, Fischer~P, Brox~T. {U-Net}: Convolutional networks for
  biomedical image segmentation. In: Medical Image Computing and
  Computer-Assisted Intervention; 2015. p. 234--241.

\bibitem{karras2022edm}
Karras~T, Aittala~M, Aila~T, et~al. Elucidating the design space of
  diffusion-based generative models. In: Advances in Neural Information
  Processing Systems; Vol.~35; 2022. p. 26565--26577.

\bibitem{li2021fno}
Li~Z, Kovachki~N, Azizzadenesheli~K, et~al. Fourier neural operator for
  parametric partial differential equations. In: International Conference on
  Learning Representations; 2021.

\bibitem{lippe2023pderefiner}
Lippe~P, Veeling~BS, Perdikaris~P, et~al. {PDE-Refiner}: Achieving accurate
  long rollouts with neural {PDE} solvers. In: Advances in Neural Information
  Processing Systems; Vol.~36; 2023. p. 67398--67433.

\end{thebibliography}


\clearpage
\appendix

\renewcommand{\thetable}{\Alph{section}\arabic{table}}
\renewcommand{\thefigure}{\Alph{section}\arabic{figure}}

\section{Experimental details and metric implementation}
\label{sec:appendix_protocol}

\setcounter{table}{0}

This appendix provides details of dataset construction, evaluation panels,
model architectures, training and sampling configurations, and metric
implementation for the experiments reported in Sections~\ref{sec:protocol} and~\ref{sec:results}.

\subsection{Datasets and splits}
\label{sec:appendix_datasets}

Table~\ref{tab:appendix_dataset_splits} summarizes the dataset dimensions, split sizes, and standard test panels. Split sizes count independent fields or trajectories, while test cases count masked reconstruction conditions. No complete field or trajectory is shared across splits.

\begin{table}[!t]
  \centering
  \scriptsize
  \setlength{\tabcolsep}{4.0pt}
  \tbl{Dataset dimensions, independent splits, and standard test panels. Split sizes count independent fields or trajectories, while test cases count masked reconstruction conditions.}{%
  \begin{tabular}{lllccc}
    \toprule
    Dataset & Reconstructed field & Complexity setting & Resolution & Train/val/test & Test cases \\
    \midrule
    Poisson & $(u,f)$ & $k_c\in\{1,2,4,6\}$ & $2\times128\times128$ & $5000/1000/1000$ & $1000$ \\
    NS & $\omega(t,x,y)$ & $\mathrm{Re}\in\{10^3,10^4,10^5\}$ & $10\times64\times64$ & $3000/300/300$ & $750$ \\
    KS & $u(t,x)$ & five $\nu$ intervals in $[0.5,1.5]$ & $1\times128\times256$ & $4096/256/512$ & $1250$ \\
    \bottomrule
  \end{tabular}}
  \label{tab:appendix_dataset_splits}
\end{table}

For Poisson, the source in Equation~\ref{eq:poisson} is generated in an
orthonormal DCT basis as
$\widehat{f}_{\mathbf{k}}=a_{\mathbf{k}}\xi_{\mathbf{k}}$, where
$\xi_{\mathbf{k}}\sim\mathcal{N}(0,1)$ and
$a_{\mathbf{k}}\propto
(1+\lVert\mathbf{k}\rVert_2^2/k_c^2)^{-1}$.
Increasing $k_c$ retains more source energy at higher wavenumbers. The zero
mode is removed, and the spectral filter is normalized to give unit expected
RMS. Each split is balanced across the four values of $k_c$.

For NS, the fixed forcing in Equation~\ref{eq:navier_stokes} is
$q(\mathbf{s})=0.1[\sin(2\pi(s_1+s_2))+\cos(2\pi(s_1+s_2))]$.
We use the released Fourier neural operator
trajectories~\cite{li2021fno} at
$\nu\in\{10^{-3},10^{-4},10^{-5}\}$, corresponding to
$\mathrm{Re}=\nu^{-1}\in\{10^3,10^4,10^5\}$. The ten retained frames
correspond to nondimensional times $t=5,\ldots,14$. For each viscosity,
trajectory indices 0--999, 1000--1099, and 1100--1199 form the training,
validation, and test splits.

For KS, we use the parameter-dependent HDF5 dataset released with
PDE-Refiner~\cite{lippe2023pderefiner} for $\nu\in[0.5,1.5]$. The original training,
validation, and test splits are retained. The domain length and time increment
vary across trajectories. We retain the first 128 time steps and all 256
spatial points without interpolation, and divide $\nu$ into five intervals
using the boundaries $\{0.5,0.7,0.9,1.1,1.3,1.5\}$.

\subsection{Observation and evaluation panels}
\label{sec:appendix_observation_panels}

Training masks for U-Net and C-EDM are generated online. For each training
example, the observation fraction is drawn uniformly from
$\{5,8,10,15,20\}\%$, after which the observed locations are sampled without
replacement. Poisson uses the same locations for $u$ and $f$, while NS and KS
use one spatial sensor set throughout each temporal window. Validation and
test masks are fixed by seed and shared across methods.

The standard Poisson test panel contains 50 independently generated fields at
each $k_c$, evaluated at all five observation fractions, yielding 1,000
reconstruction conditions. The NS and KS panels contain 50 trajectories per
parameter regime. Evaluating each trajectory at all five observation
fractions yields 750 NS and 1,250 KS reconstruction conditions.

Additional observation-fraction panels evaluate 3\% and 12\% masks on 200
Poisson fields, 150 NS trajectories, and 250 KS trajectories. The 3\%
setting extrapolates below the training range, whereas 12\% interpolates
within it. Poisson source-frequency generalization is evaluated using 50 new
fields at each of $k_c=3.5$ and $k_c=8$ across all five standard observation
fractions. Mask-distribution robustness is evaluated on 200 Poisson fields
and 150 NS trajectories at 5\% observations. Each shifted mask is paired with
a uniform mask on the same field or trajectory. KS is excluded because its
one-dimensional spatial sensors have no direct counterpart to these
two-dimensional mask shifts. The shifted mask distributions are defined in
Appendix~\ref{sec:appendix_robustness}.

\subsection{Architecture, training, and sampling configurations}
\label{sec:appendix_training}

All methods use architecture-matched U-Net backbones with approximately
14.3 million parameters. Each backbone has base width 64, channel
multipliers $(1,2,2,4)$, two residual blocks per level, attention at
resolution 16, and dropout 0.1. Optimization uses Adam with an initial
learning rate of $2\times10^{-4}$, a batch size of 32, automatic mixed
precision, gradient clipping at norm 1, and an EMA half-life of
$50{,}000$ training fields. The learning rate is warmed up for five epochs
and decayed to $2\times10^{-5}$ by epoch 750. All models are trained with
seed 0.

For both EDM formulations, $\sigma_{\mathrm{data}}=1$ and the training noise
levels follow
$\log\sigma\sim\mathcal{N}(-0.5,1.2^2)$. Sampling uses the Karras schedule
over $\sigma\in[0.002,80]$ with $\rho=7$. The Poisson variables are
standardized separately by channel, whereas NS and KS use one global mean and
standard deviation. All normalization statistics are computed from the
training split.

Table~\ref{tab:appendix_training_protocol} reports the training
budgets, EMA checkpoints selected according to validation performance, and
test-time sampling configurations. For G-EDM, the guidance multiplier is
fixed at $r_i=0.1$ during the final 20\% of sampling steps.
Appendix~\ref{sec:appendix_guidance} details the selection of the sampling
steps and guidance weights. All reported EDM ensembles use
$N_{\mathrm{ens}}=32$, with the corresponding ensemble-size analysis
provided in Appendix~\ref{sec:appendix_ensemble_size}.

\begin{table}[!t]
  \centering
  \scriptsize
  \setlength{\tabcolsep}{4.0pt}
  \tbl{Training epochs, EMA checkpoints used for evaluation, and sampling configurations. U, C, and G denote U-Net, C-EDM, and G-EDM. The Poisson guidance weights are reported as $(\zeta_u,\zeta_f)$.}{%
  \begin{tabular}{lcccc}
    \toprule
    Dataset & Training epochs U/C/G & EMA epoch U/C/G
    & C-EDM $N$ & G-EDM $(N,\zeta)$ \\
    \midrule
    Poisson & $1000/1000/1000$ & $1000/1000/1000$
    & $40$ & $400,\ (10,20)$ \\
    NS & $1500/1500/1500$ & $1500/1470/1360$
    & $40$ & $1600,\ 5$ \\
    KS & $1200/1200/1000$ & $1200/1040/960$
    & $20$ & $1600,\ 70$ \\
    \bottomrule
  \end{tabular}}
  \label{tab:appendix_training_protocol}
\end{table}

\subsection{Metric implementation}
\label{sec:appendix_metrics}

Before computing reconstruction and spectral metrics, predicted values at
observed locations are replaced with the corresponding measurements according
to
\begin{equation}
  \widehat{\mathbf{x}}_0^{\mathrm{eval}}
  =
  \mathbf{y}
  +
  (\mathbf{1}-\mathbf{M})
  \odot
  \widehat{\mathbf{x}}_0.
  \label{eq:measurement_restoration}
\end{equation}
Reconstruction and spectral errors are computed in physical units for each
variable before averaging across test cases. Probabilistic metrics are
computed using standardized values only at unobserved locations.

For method $m$ and shifted mask distribution $g$, mask sensitivity is computed
for each paired test case as
$\Delta\varepsilon_{m,g}
=\varepsilon_{m,g}-\varepsilon_{m,\mathrm{uniform}}$.
Both masks contain 5\% observations, and the reported value is the mean paired
difference across test cases. For any scalar reconstruction metric $d$, the
relative reduction from the mean memberwise statistic to the ensemble-mean
statistic is calculated as
$100(d_{\mathrm{S}}-d_{\mathrm{E}})/d_{\mathrm{S}}$.

The transform $\mathcal{T}$ in
Equation~\ref{eq:bandwise_spectral_error} is benchmark-specific. Poisson $u$
uses a type-I discrete sine transform of the interior field, consistent with
its homogeneous Dirichlet boundary condition, while Poisson $f$ uses a type-II
discrete cosine transform. For NS and KS, the temporal dimension is retained,
and $\mathcal{T}$ is applied independently over space at each time step. NS
uses a two-dimensional spatial Fourier transform, while KS uses a
one-dimensional spatial Fourier transform and retains modes
$1\leq k\leq60$. The frequency-band norms then accumulate the coefficient
errors over time. For the two-dimensional fields, radial power is obtained by
summing squared coefficient magnitudes within each integer-wavenumber shell.
The dataset-specific frequency bands are shown in Appendix
Figure~\ref{fig:truth_spectra}.

Because the frequency bands form a disjoint partition of $\mathcal{K}$ and
share the same full-spectrum denominator,
Equation~\ref{eq:bandwise_spectral_error} satisfies
\begin{equation}
  \left(\varepsilon_{\mathrm{all}}^{\mathrm{spec}}\right)^2
  =
  \sum_b
  \left(\varepsilon_b^{\mathrm{spec}}\right)^2.
  \label{eq:spectral_error_partition}
\end{equation}
Each band therefore retains its contribution to the full-spectrum coefficient
error. The log-power metric instead averages wavenumberwise discrepancies
after power normalization and does not weight frequency bands by their
contribution to the total spectral energy.

For a standardized scalar target $x$ at an unobserved location and ensemble
predictions
$\{\widehat{x}^{(m)}\}_{m=1}^{N_{\mathrm{ens}}}$, the empirical continuous
ranked probability score is
\begin{equation}
  \mathrm{CRPS}
  =
  \frac{1}{N_{\mathrm{ens}}}
  \sum_{m=1}^{N_{\mathrm{ens}}}
  \left|\widehat{x}^{(m)}-x\right|
  -
  \frac{1}{2N_{\mathrm{ens}}^2}
  \sum_{m=1}^{N_{\mathrm{ens}}}
  \sum_{n=1}^{N_{\mathrm{ens}}}
  \left|\widehat{x}^{(m)}-\widehat{x}^{(n)}\right|.
  \label{eq:empirical_crps}
\end{equation}
CRPS is averaged over unobserved entries and then over test cases. The central
90\% prediction interval is formed from the empirical 0.05 and 0.95
quantiles. Its coverage is the fraction of unobserved targets within the
interval, and its width is the mean difference between the two quantiles.

Per-output inference latency is measured on an NVIDIA RTX 4090 with batch size
1 and mixed precision after warming up each inference graph. Measurements
exclude data loading, metric computation, and file I/O. Neural function
evaluations count denoiser calls. For G-EDM, observation-gradient corrections
are counted separately as gradient evaluations. The 95\% confidence intervals
shown in the figures are computed by paired bootstrap resampling, using an
independent field or trajectory as the sampling unit.

\section{Spectral results and metric sensitivity}

\label{sec:appendix_spectral_results}

\setcounter{table}{0}
\setcounter{figure}{0}

\begin{figure}[!t]
  \centering
  \includegraphics[width=\textwidth]{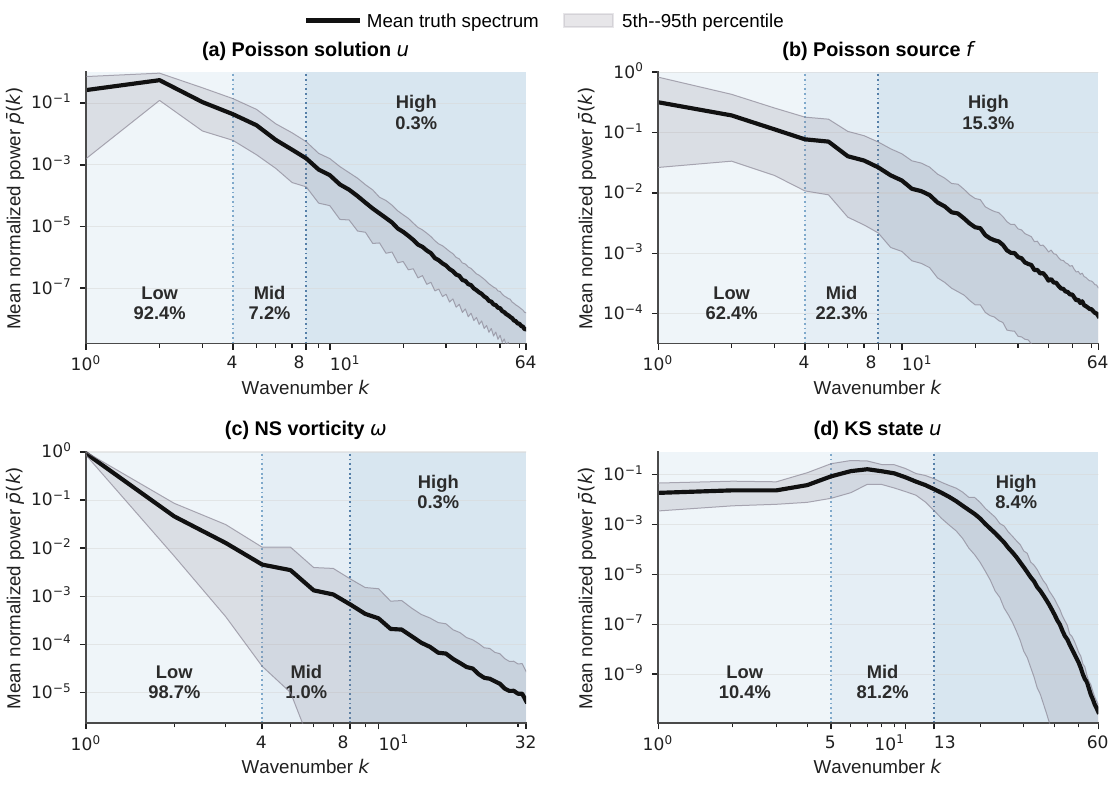}
  \caption{Normalized ground-truth power spectra partitioned into dataset-specific low-, mid-, and high-frequency bands. Lines show test-set means, shaded regions indicate the 5th--95th percentile ranges, and annotations report the mean energy fraction in each band.}
  \label{fig:truth_spectra}
\end{figure}

Dataset-specific frequency bands were selected from the ground-truth normalized power spectra shown in Figure~\ref{fig:truth_spectra}. For Poisson and NS, the low-, mid-, and high-frequency bands are $0\leq k<4$, $4\leq k<8$, and $k\geq8$. For KS, the corresponding bands are $1\leq k<5$, $5\leq k<13$, and $13\leq k\leq60$. Poisson $u$ and NS $\omega$ are dominated by low-frequency energy, whereas Poisson $f$ and KS $u$ contain larger fractions of energy in the mid- and high-frequency bands. Tables~\ref{tab:appendix_poisson_spectrum}--\ref{tab:appendix_ks_spectrum} report $D_b^{\log P}$ for all complexity regimes and frequency bands, including the high-frequency results summarized in Table~\ref{tab:ensemble_spectrum_effect}.

\FloatBarrier

\begin{table}[!t]
  \centering
  \scriptsize
  \setlength{\tabcolsep}{3.4pt}
  \tbl{Bandwise log-power error $D_b^{\log P}$ for Poisson $u$ and $f$ across $k_c$, averaged over test fields and observation fractions. Best and second-best results are shown in bold and underlined, respectively.}{%
  \begin{tabular}{lllccccc}
    \toprule
    Field & Regime & Wavenumber range & U-Net & C-EDM (E) & C-EDM (S) & G-EDM (E) & G-EDM (S) \\
    \midrule
    Poisson $u$ & $k_c=1$ & $0\leq k<4$ & $0.00378$ & $0.00973$ & $0.00815$ & $\mathbf{0.00067}$ & $0.00112$ \\
     &  & $4\leq k<8$ & $0.00124$ & $\mathbf{0.00111}$ & $0.00163$ & $0.00139$ & $0.00167$ \\
     &  & $k\geq8$ & $0.2470$ & $\mathbf{0.1268}$ & $0.1762$ & $0.3032$ & $0.4145$ \\
    \addlinespace[1.5pt]
    Poisson $u$ & $k_c=2$ & $0\leq k<4$ & $0.00060$ & $0.00117$ & $0.00230$ & $\mathbf{0.00015}$ & $0.00021$ \\
     &  & $4\leq k<8$ & $\mathbf{0.00094}$ & $0.00101$ & $0.00146$ & $0.00120$ & $0.00155$ \\
     &  & $k\geq8$ & $0.2149$ & $\mathbf{0.0963}$ & $0.1493$ & $0.2818$ & $0.3860$ \\
    \addlinespace[1.5pt]
    Poisson $u$ & $k_c=4$ & $0\leq k<4$ & $0.00054$ & $0.00115$ & $0.00240$ & $\mathbf{0.00032}$ & $0.00037$ \\
     &  & $4\leq k<8$ & $\mathbf{0.00076}$ & $0.00077$ & $0.00121$ & $0.00119$ & $0.00144$ \\
     &  & $k\geq8$ & $0.1753$ & $\mathbf{0.0730}$ & $0.1286$ & $0.1996$ & $0.2981$ \\
    \addlinespace[1.5pt]
    Poisson $u$ & $k_c=6$ & $0\leq k<4$ & $0.00093$ & $0.00174$ & $0.00346$ & $\mathbf{0.00079}$ & $0.00090$ \\
     &  & $4\leq k<8$ & $\mathbf{0.00082}$ & $0.00084$ & $0.00128$ & $0.00211$ & $0.00254$ \\
     &  & $k\geq8$ & $0.1678$ & $\mathbf{0.0739}$ & $0.1227$ & $0.2147$ & $0.3171$ \\
    \addlinespace[1.5pt]
    Poisson $f$ & $k_c=1$ & $0\leq k<4$ & $\mathbf{0.00123}$ & $0.00128$ & $0.00155$ & $0.00156$ & $0.00170$ \\
     &  & $4\leq k<8$ & $0.00412$ & $\mathbf{0.00405}$ & $0.00591$ & $0.00620$ & $0.00689$ \\
     &  & $k\geq8$ & $0.3586$ & $0.3348$ & $\mathbf{0.0630}$ & $0.2235$ & $0.1509$ \\
    \addlinespace[1.5pt]
    Poisson $f$ & $k_c=2$ & $0\leq k<4$ & $0.00206$ & $0.00214$ & $\mathbf{0.00195}$ & $0.00229$ & $0.00246$ \\
     &  & $4\leq k<8$ & $\mathbf{0.00360}$ & $0.00392$ & $0.00516$ & $0.00607$ & $0.00692$ \\
     &  & $k\geq8$ & $0.3356$ & $0.3170$ & $\mathbf{0.0600}$ & $0.2460$ & $0.1447$ \\
    \addlinespace[1.5pt]
    Poisson $f$ & $k_c=4$ & $0\leq k<4$ & $0.00540$ & $0.00599$ & $\mathbf{0.00380}$ & $0.00799$ & $0.00872$ \\
     &  & $4\leq k<8$ & $\mathbf{0.00536}$ & $0.00618$ & $0.00542$ & $0.00820$ & $0.00929$ \\
     &  & $k\geq8$ & $0.2940$ & $0.2918$ & $\mathbf{0.0570}$ & $0.2392$ & $0.0926$ \\
    \addlinespace[1.5pt]
    Poisson $f$ & $k_c=6$ & $0\leq k<4$ & $0.0110$ & $0.0129$ & $\mathbf{0.00733}$ & $0.0145$ & $0.0133$ \\
     &  & $4\leq k<8$ & $0.0107$ & $0.0124$ & $\mathbf{0.00749}$ & $0.0133$ & $0.0127$ \\
     &  & $k\geq8$ & $0.2703$ & $0.2778$ & $\mathbf{0.0557}$ & $0.2256$ & $0.0704$ \\
    \bottomrule
  \end{tabular}}
  \label{tab:appendix_poisson_spectrum}
\end{table}

\begin{table}[!t]
  \centering
  \scriptsize
  \setlength{\tabcolsep}{3.4pt}
  \tbl{Bandwise log-power error $D_b^{\log P}$ for NS $\omega$ across Reynolds numbers, averaged over test trajectories and observation fractions. Best and second-best results are shown in bold and underlined, respectively.}{%
  \begin{tabular}{lllccccc}
    \toprule
    Field & Regime & Wavenumber range & U-Net & C-EDM (E) & C-EDM (S) & G-EDM (E) & G-EDM (S) \\
    \midrule
    NS $\omega$ & $Re=10^{3}$ & $0\leq k<4$ & $0.00240$ & $0.00241$ & $0.00435$ & $\mathbf{0.00045}$ & $0.00064$ \\
     &  & $4\leq k<8$ & $0.1961$ & $0.1229$ & $0.3174$ & $\mathbf{0.1155}$ & $0.1636$ \\
     &  & $k\geq8$ & $6.2367$ & $\mathbf{6.1541}$ & $6.8390$ & $6.2545$ & $6.4540$ \\
    \addlinespace[1.5pt]
    NS $\omega$ & $Re=10^{4}$ & $0\leq k<4$ & $0.00146$ & $0.00188$ & $0.00267$ & $\mathbf{0.00054}$ & $0.00069$ \\
     &  & $4\leq k<8$ & $0.00937$ & $0.0132$ & $0.0119$ & $\mathbf{0.00320}$ & $0.00346$ \\
     &  & $k\geq8$ & $0.5889$ & $0.6137$ & $0.7922$ & $\mathbf{0.5128}$ & $0.6448$ \\
    \addlinespace[1.5pt]
    NS $\omega$ & $Re=10^{5}$ & $0\leq k<4$ & $0.00286$ & $0.00306$ & $0.00382$ & $\mathbf{0.00263}$ & $0.00303$ \\
     &  & $4\leq k<8$ & $0.0165$ & $0.0207$ & $0.0177$ & $\mathbf{0.0105}$ & $0.0111$ \\
     &  & $k\geq8$ & $0.1674$ & $0.1646$ & $0.0625$ & $0.1204$ & $\mathbf{0.0371}$ \\
    \bottomrule
  \end{tabular}}
  \label{tab:appendix_ns_spectrum}
\end{table}

\begin{table}[!t]
  \centering
  \scriptsize
  \setlength{\tabcolsep}{3.4pt}
  \tbl{Bandwise log-power error $D_b^{\log P}$ for KS $u$ across viscosity intervals, averaged over test trajectories and observation fractions. Best and second-best results are shown in bold and underlined, respectively.}{%
  \begin{tabular}{lllccccc}
    \toprule
    Field & Regime & Wavenumber range & U-Net & C-EDM (E) & C-EDM (S) & G-EDM (E) & G-EDM (S) \\
    \midrule
    KS $u$ & $\nu\in[0.5,0.7)$ & $1\leq k<5$ & $\mathbf{0.00636}$ & $0.0104$ & $0.0140$ & $0.0178$ & $0.0257$ \\
     &  & $5\leq k<13$ & $\mathbf{0.00475}$ & $0.00778$ & $0.0113$ & $0.0140$ & $0.0197$ \\
     &  & $13\leq k\leq60$ & $0.9301$ & $\mathbf{0.5797}$ & $0.6930$ & $0.9379$ & $0.9988$ \\
    \addlinespace[1.5pt]
    KS $u$ & $\nu\in[0.7,0.9)$ & $1\leq k<5$ & $\mathbf{0.00561}$ & $0.00685$ & $0.00933$ & $0.0166$ & $0.0219$ \\
     &  & $5\leq k<13$ & $\mathbf{0.00394}$ & $0.00509$ & $0.00715$ & $0.0106$ & $0.0139$ \\
     &  & $13\leq k\leq60$ & $1.5144$ & $\mathbf{0.9481}$ & $1.1226$ & $1.6035$ & $1.6682$ \\
    \addlinespace[1.5pt]
    KS $u$ & $\nu\in[0.9,1.1)$ & $1\leq k<5$ & $0.00571$ & $\mathbf{0.00566}$ & $0.00869$ & $0.0142$ & $0.0193$ \\
     &  & $5\leq k<13$ & $\mathbf{0.00360}$ & $0.00537$ & $0.00708$ & $0.00870$ & $0.0124$ \\
     &  & $13\leq k\leq60$ & $2.2203$ & $\mathbf{1.4960}$ & $1.7430$ & $2.3292$ & $2.4199$ \\
    \addlinespace[1.5pt]
    KS $u$ & $\nu\in[1.1,1.3)$ & $1\leq k<5$ & $0.00629$ & $\mathbf{0.00507}$ & $0.00741$ & $0.0155$ & $0.0208$ \\
     &  & $5\leq k<13$ & $0.00412$ & $\mathbf{0.00406}$ & $0.00623$ & $0.00790$ & $0.0118$ \\
     &  & $13\leq k\leq60$ & $2.7283$ & $\mathbf{1.9064}$ & $2.1920$ & $2.9233$ & $2.9899$ \\
    \addlinespace[1.5pt]
    KS $u$ & $\nu\in[1.3,1.5)$ & $1\leq k<5$ & $0.00608$ & $\mathbf{0.00323}$ & $0.00548$ & $0.0142$ & $0.0191$ \\
     &  & $5\leq k<13$ & $0.00494$ & $\mathbf{0.00261}$ & $0.00452$ & $0.0101$ & $0.0132$ \\
     &  & $13\leq k\leq60$ & $3.3289$ & $\mathbf{2.3433}$ & $2.6807$ & $3.5621$ & $3.5890$ \\
    \bottomrule
  \end{tabular}}
  \label{tab:appendix_ks_spectrum}
\end{table}

\FloatBarrier

\subsection{Sensitivity to the spectral metric}

The high-frequency analysis in
Section~\ref{sec:ensemble_spectral_effects} uses the log-power error,
which measures discrepancies at each wavenumber on a logarithmic scale
and remains sensitive to modes with low normalized power. To examine
whether the method rankings depend on this emphasis, we also compare
unit-total-power spectra using the relative $\ell_2$ distance
\begin{equation}
  R_{\widetilde P}
  =
  \frac{
    \left\|\widetilde P_{\mathrm{pred}}
    -\widetilde P_{\mathrm{true}}\right\|_2
  }{
    \left\|\widetilde P_{\mathrm{true}}\right\|_2
  }.
  \label{eq:normalized_power_distance}
\end{equation}
The denominator is a global norm of the ground-truth spectrum rather
than a pointwise normalization at each wavenumber. Consequently,
$R_{\widetilde P}$ is driven mainly by absolute discrepancies in
energetic modes and is less sensitive to the low-power spectral tail
than $D_b^{\log P}$.

\begin{table}[!t]
  \centering
  \small
  \setlength{\tabcolsep}{4.2pt}
  \tbl{Mean relative $\ell_2$ distance $R_{\widetilde P}$ between normalized power spectra across test cases, field-complexity settings, and observation fractions. The lowest and second-lowest values are shown in bold and underlined, respectively.}{%
  \begin{tabular}{lccccc}
    \toprule
    Field & U-Net & C-EDM (E) & C-EDM (S) & G-EDM (E) & G-EDM (S) \\
    \midrule
    Poisson $u$ & $3.586\!\times\!10^{-4}$ & $6.811\!\times\!10^{-4}$ & $1.220\!\times\!10^{-3}$ & $\mathbf{2.786\!\times\!10^{-4}}$ & $\underline{3.454\!\times\!10^{-4}}$ \\
    Poisson $f$ & $\underline{1.200\!\times\!10^{-2}}$ & $1.362\!\times\!10^{-2}$ & $\mathbf{8.819\!\times\!10^{-3}}$ & $1.584\!\times\!10^{-2}$ & $1.578\!\times\!10^{-2}$ \\
    NS $\omega$ & $1.464\!\times\!10^{-3}$ & $1.669\!\times\!10^{-3}$ & $1.713\!\times\!10^{-3}$ & $\mathbf{9.437\!\times\!10^{-4}}$ & $\underline{1.034\!\times\!10^{-3}}$ \\
    KS $u$ & $\mathbf{1.548\!\times\!10^{-2}}$ & $\underline{1.846\!\times\!10^{-2}}$ & $2.704\!\times\!10^{-2}$ & $3.589\!\times\!10^{-2}$ & $5.046\!\times\!10^{-2}$ \\
    \bottomrule
  \end{tabular}}
  \label{tab:appendix_normalized_power_spectrum}
\end{table}

Table~\ref{tab:appendix_normalized_power_spectrum} reports
$R_{\widetilde P}$ averaged over each test set. G-EDM (E) achieved the
lowest distance for Poisson $u$ and NS $\omega$, C-EDM (S) for Poisson
$f$, and U-Net for KS $u$. This ordering differs from the high-frequency
log-power comparison, showing that recovery of energetic modes and
recovery of the low-power spectral tail are distinct aspects of spectral
reconstruction.

\FloatBarrier

\section{Observation-process robustness and source-frequency generalization}
\label{sec:appendix_robustness}

\setcounter{table}{0}
\setcounter{figure}{0}

\begin{figure}[!t]
  \centering
  \includegraphics[width=\textwidth]{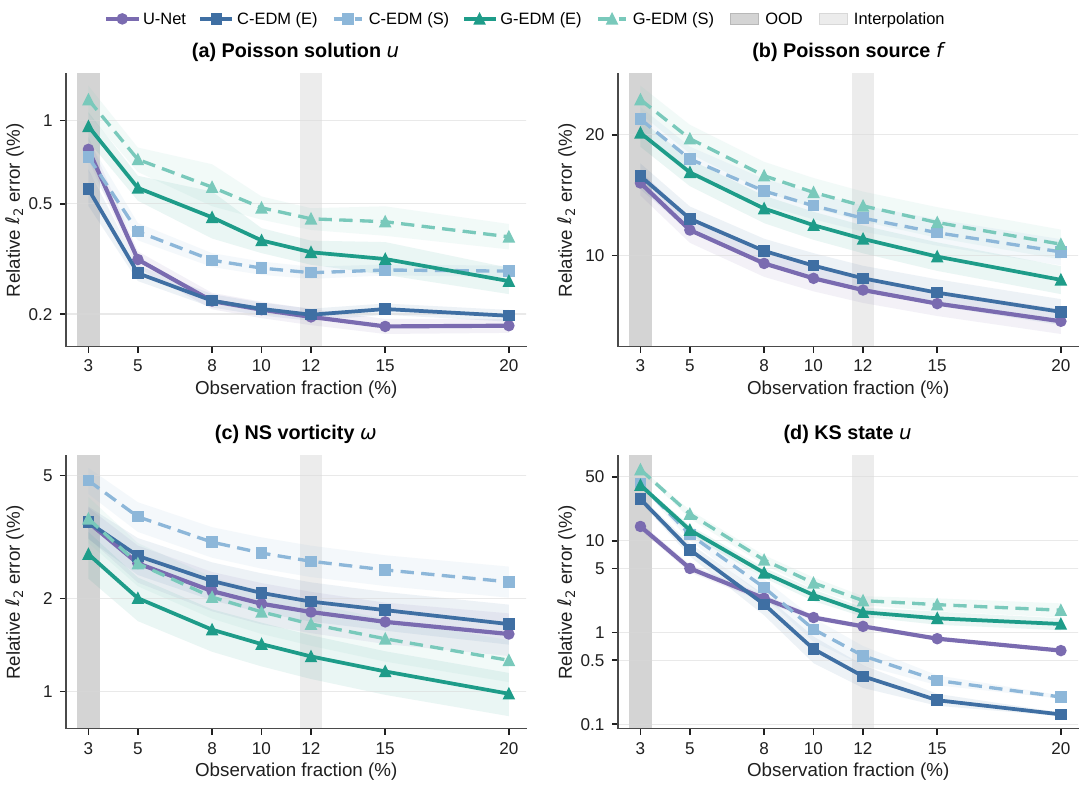}
  \caption{Full-field relative $\ell_2$ error versus observation fraction for Poisson $u$ and $f$, NS $\omega$, and KS $u$. The 3\% setting extrapolates below the training range, while the 12\% setting interpolates between training fractions. Bands show paired-bootstrap 95\% confidence intervals.}
  \label{fig:appendix_observation_fraction}
\end{figure}

Figure~\ref{fig:appendix_observation_fraction} extends the five observation fractions used during training with an extrapolation below the training range at 3\% and an interpolation at 12\%. Full-field relative $\ell_2$ error generally decreased as the observation fraction increased. Both additional fractions followed the neighboring trends without changing the method rankings.

\begin{figure}[!t]
  \centering
  \includegraphics[width=0.99\textwidth]{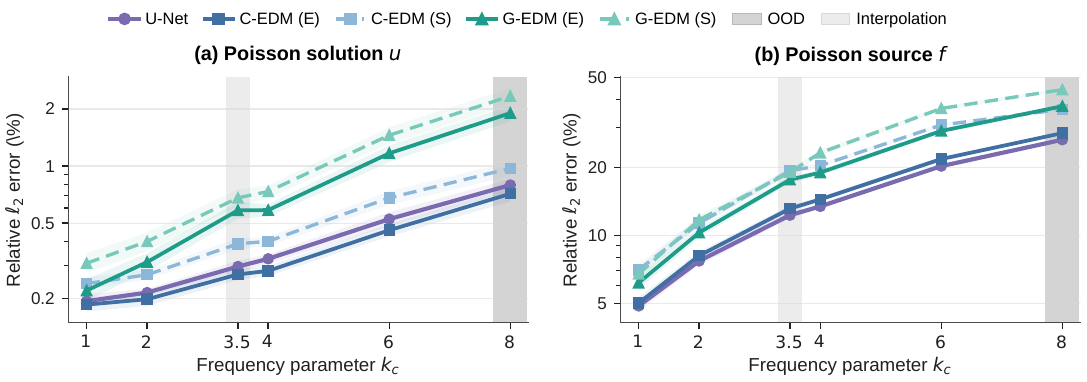}
  \caption{Full-field relative $\ell_2$ error versus the Poisson source-frequency parameter $k_c$ at 5\% observations. The $k_c=3.5$ case interpolates between training frequencies, while $k_c=8$ extrapolates beyond the training range. Bands show bootstrap 95\% confidence intervals.}
  \label{fig:appendix_poisson_frequency_generalization}
\end{figure}

Figure~\ref{fig:appendix_poisson_frequency_generalization} evaluates Poisson source-frequency generalization at 5\% observations. The $k_c=3.5$ interpolation followed the trend between neighboring training frequencies, while the $k_c=8$ extrapolation continued the increase in error beyond the upper training boundary. C-EDM (E) retained the lowest error for Poisson $u$, and U-Net retained the lowest error for Poisson $f$.

The mask distributions shown in Figure~\ref{fig:mask_examples} were evaluated on paired test fields or trajectories at a fixed 5\% observation count. Each shifted mask was compared with a uniform mask on the same test case. Gaussian--uniform masks placed half of the sensors in an isotropic Gaussian cluster with a random center and a standard deviation of 0.1 on the unit domain. The remaining sensors were sampled uniformly. Directional masks placed 75\% of the sensors in one half of the domain, with the four orientations balanced across test cases. Patch-missing masks excluded a randomly positioned square with a side length equal to 25\% of the domain and sampled the remaining sensors uniformly. For NS, the same mask was applied to every temporal frame, with periodic wrapping of the missing patch.

\begin{figure}[!t]
  \centering
  \includegraphics[width=\textwidth]{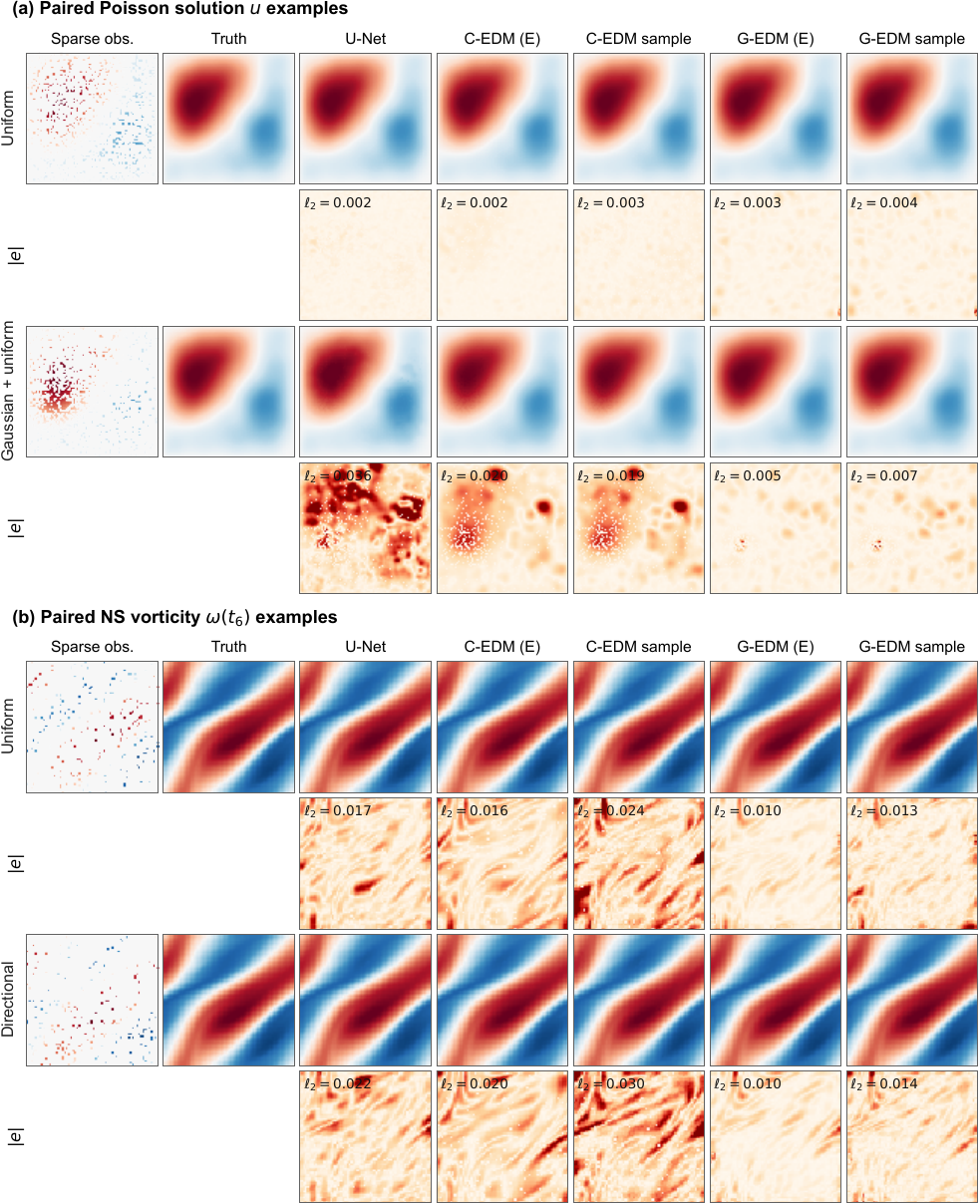}
  \caption{Paired reconstructions and absolute-error maps at 5\% observations. Panel (a) compares Poisson $u$ under uniform and Gaussian--uniform masks, and panel (b) compares NS $\omega(t_6)$ under uniform and directional masks. Each reconstruction row is followed by the corresponding absolute-error map. Annotations report the full-field relative $\ell_2$ error.}
  \label{fig:appendix_mask_cases}
\end{figure}

Figure~\ref{fig:appendix_mask_cases} presents additional paired reconstructions for Poisson $u$ under Gaussian--uniform masks and NS $\omega(t_6)$ under directional masks. 
Table~\ref{tab:appendix_mask_robustness} reports the corresponding mean full-field relative $\ell_2$ errors under the uniform baseline and the three shifted mask distributions, complementing the baseline-relative increases in Figure~\ref{fig:mask_robustness}.

\clearpage

\begin{table}[!t]
  \centering
  \small
  \setlength{\tabcolsep}{4.0pt}
  \tbl{Mean full-field relative $\ell_2$ error (\%) under the uniform baseline and three shifted mask distributions at a fixed 5\% observation count, evaluated on paired test fields or trajectories. The lowest and second-lowest values are shown in bold and underlined, respectively.}{%
  \begin{tabular}{llccccc}
    \toprule
    Field & Mask & U-Net & C-EDM (E) & C-EDM (S)
      & G-EDM (E) & G-EDM (S) \\
    \midrule
    Poisson $u$ & Uniform
      & \underline{0.315} & \textbf{0.281} & 0.397 & 0.572 & 0.724 \\
      & Gaussian + uniform
      & 3.096 & 1.934 & 2.091 & \textbf{1.140} & \underline{1.454} \\
      & Directional bias
      & \underline{0.782} & \textbf{0.632} & 0.806 & 0.874 & 1.114 \\
      & Patch missing
      & 4.428 & 8.007 & 8.337 & \textbf{1.068} & \underline{1.355} \\
    \addlinespace[2pt]
    Poisson $f$ & Uniform
      & \textbf{11.573} & \underline{12.347} & 17.400 & 16.118 & 19.554 \\
      & Gaussian + uniform
      & 25.728 & \underline{20.815} & 26.985 & \textbf{20.771} & 26.151 \\
      & Directional bias
      & \textbf{13.950} & \underline{14.877} & 20.645 & 18.303 & 22.658 \\
      & Patch missing
      & 24.298 & 37.023 & 41.695 & \textbf{17.133} & \underline{20.936} \\
    \addlinespace[2pt]
    NS $\omega$ & Uniform
      & 2.599 & 2.748 & 3.681 & \textbf{2.005} & \underline{2.597} \\
      & Gaussian + uniform
      & 3.672 & 3.615 & 4.843 & \textbf{2.621} & \underline{3.459} \\
      & Directional bias
      & 3.116 & 3.213 & 4.356 & \textbf{2.222} & \underline{2.901} \\
      & Patch missing
      & 2.802 & 2.928 & 3.966 & \textbf{2.099} & \underline{2.720} \\
    \bottomrule
  \end{tabular}}
  \label{tab:appendix_mask_robustness}
\end{table}

\section{Additional spatial uncertainty examples}
\label{sec:appendix_distributional_results}

\begin{figure}[!t]
  \centering
  \includegraphics[width=\textwidth]{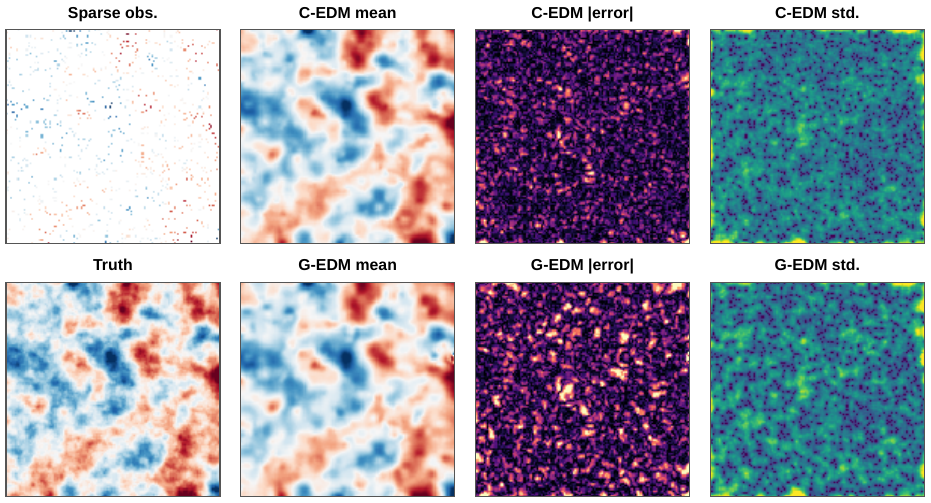}
  \caption{Spatial uncertainty for a representative Poisson $f$ field at $k_c=6$ with 5\% uniformly sampled observations. Panels show observations, ground truth, ensemble means, absolute errors, and standard deviations for both EDM formulations ($N_{\mathrm{ens}}=32$).}
  \label{fig:appendix_uncertainty_poisson_f}
\end{figure}

\begin{figure}[!t]
  \centering
  \includegraphics[width=\textwidth]{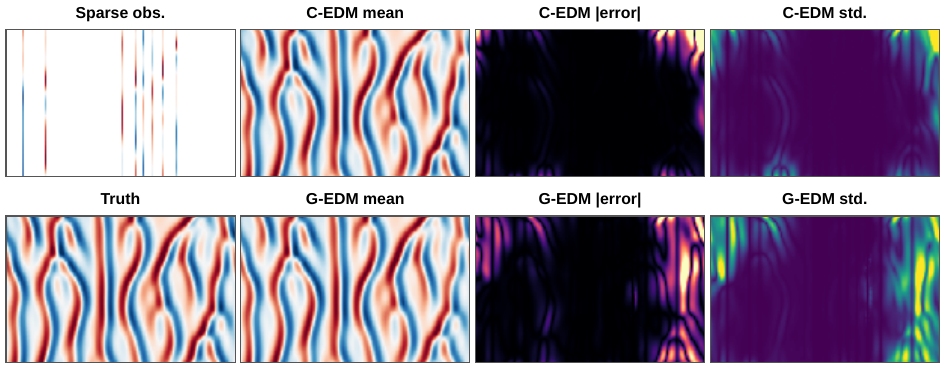}
  \caption{Spatial uncertainty for a representative KS $u(t,x)$ trajectory with 5\% uniformly sampled observations. Panels show observations, ground truth, ensemble means, absolute errors, and standard deviations for both EDM formulations ($N_{\mathrm{ens}}=32$).}
  \label{fig:appendix_uncertainty_ks}
\end{figure}

Figures~\ref{fig:appendix_uncertainty_poisson_f} and~
\ref{fig:appendix_uncertainty_ks}extend the spatial uncertainty analysis in Figure~\ref{fig:uncertainty_maps} to Poisson $f$ and KS $u$ underthe same 5\% uniformly sampled observation protocol.  For each EDMformulation, the figures compare ensemble standard deviation withabsolute reconstruction error. Both quantities are normalized by theRMS amplitude of the corresponding ground-truth field. In bothexamples, regions of larger ensemble spread broadly coincided withregions of larger reconstruction error.

\FloatBarrier

\section{Prior-guided sampling parameter selection}
\label{sec:appendix_guidance}

\setcounter{table}{0}

G-EDM sampling parameters were selected on validation data. $N$ denotes the number of sampling steps, and $\zeta$ denotes the per-step observation-guidance weight. Poisson uses separate weights $(\zeta_u,\zeta_f)$ for the solution and source fields. The selection score $J_{\mathrm{val}}$ is the macro-average ensemble-mean relative $\ell_2$ error over unobserved locations. For Poisson, the score assigns equal weight to the two reconstructed fields. Candidate configurations with an observed-location normalized RMSE above 0.1 were discarded.

Parameter selection followed two stages. Each configuration was first evaluated using one generated sample per case on a balanced screening panel. Promising configurations were then evaluated using four samples per case on a larger, disjoint confirmation panel. The screening and confirmation panels contained 40 and 200 cases for Poisson, 30 and 150 for NS, and 25 and 125 for KS. Validation masks and sampling seeds were fixed across configurations. The final-20\% guidance multiplier was also fixed at $r_i=0.1$. Tables~
\ref{tab:appendix_guidance_poisson} and~
\ref{tab:appendix_guidance_ns_ks} report the evaluated combinations of sampling steps and guidance weights.

\begin{table}[!t]
  \centering
  \setlength{\tabcolsep}{7pt}
  \renewcommand{\arraystretch}{0.90}
  \tbl{Validation results for selecting the G-EDM sampling steps and guidance weights for Poisson. The guidance weights are $(\zeta_u,\zeta_f)$. Boldface indicates the lowest $J_{\mathrm{val}}$ within each selection stage, and $\dagger$ marks the configuration used for test evaluation.}{%
  \begin{tabular*}{0.78\textwidth}{@{\extracolsep{\fill}}lrrr@{}}
    \toprule
    $N$ & Weight & $J_{\mathrm{val}}\downarrow$ & Obs. NRMSE $\downarrow$ \\
    \midrule
    \multicolumn{4}{l}{\textit{Screen}} \\
    400 & $(5,5)$   & 0.08502 & 0.00645 \\
    400 & $(5,10)$  & 0.07906 & 0.01199 \\
    400 & $(5,20)$  & 0.07659 & 0.02548 \\
    400 & $(10,10)$ & 0.07866 & 0.01497 \\
    400 & $(10,20)$ & \textbf{0.07653} & 0.02622 \\
    400 & $(10,40)$ & 0.07836 & 0.04348 \\
    400 & $(20,20)$ & 0.07890 & 0.03153 \\
    400 & $(20,40)$ & 0.11623 & 0.05111 \\
    400 & $(20,80)$ & 0.16804 & 0.09568 \\
    \addlinespace[2pt]
    \multicolumn{4}{l}{\textit{Confirm}} \\
    40  & $(25,50)$ & 0.09242 & 0.06528 \\
    100 & $(20,40)$ & 0.07019 & 0.04473 \\
    200 & $(10,20)$ & 0.06602 & 0.02493 \\
    $400^{\dagger}$ & $(10,20)$ & \textbf{0.06570} & 0.02579 \\
    800 & $(9,19)$ & 0.06648 & 0.02369 \\
    \bottomrule
  \end{tabular*}}
  \label{tab:appendix_guidance_poisson}
\end{table}

\begin{table}[!t]
  \centering
  \setlength{\tabcolsep}{3.5pt}
  \renewcommand{\arraystretch}{0.90}
  \tbl{Validation results for selecting the G-EDM sampling steps and guidance weights for NS and KS. Both benchmarks use a single guidance weight $\zeta$. Boldface indicates the lowest $J_{\mathrm{val}}$ within each selection stage, and $\dagger$ marks the configuration used for test evaluation.}{%
  \begin{minipage}[t]{0.465\textwidth}
    \vspace{0pt}
    \centering
    \begin{tabular*}{\linewidth}[t]{@{\extracolsep{\fill}}lrrr@{}}
      \toprule
      \multicolumn{4}{l}{\textit{(a) NS}} \\
      $N$ & Weight & $J_{\mathrm{val}}\downarrow$ & Obs. NRMSE $\downarrow$ \\
      \midrule
      \multicolumn{4}{l}{\textit{Screen}} \\
      400   & 20 & 0.02207 & 0.01615 \\
      600   & 13 & 0.02083 & 0.01042 \\
      800   & 10 & 0.02032 & 0.00786 \\
      1,200 & 7  & 0.01983 & 0.00540 \\
      1,600 & 5  & 0.01959 & 0.00399 \\
      2,000 & 4  & \textbf{0.01946} & 0.00309 \\
      \addlinespace[2pt]
      \multicolumn{4}{l}{\textit{Confirm}} \\
      40    & 50 & 0.05067 & 0.04081 \\
      100   & 40 & 0.02457 & 0.03316 \\
      200   & 20 & 0.01975 & 0.01653 \\
      400   & 10 & 0.01809 & 0.00780 \\
      1,200 & 7  & 0.01571 & 0.00524 \\
      $1{,}600^{\dagger}$ & 5 & 0.01536 & 0.00388 \\
      2,000 & 4 & \textbf{0.01517} & 0.00309 \\
      \bottomrule
    \end{tabular*}
  \end{minipage}\hspace{0.035\textwidth}%
  \begin{minipage}[t]{0.465\textwidth}
    \vspace{0pt}
    \centering
    \begin{tabular*}{\linewidth}[t]{@{\extracolsep{\fill}}lrrr@{}}
      \toprule
      \multicolumn{4}{l}{\textit{(b) KS}} \\
      $N$ & Weight & $J_{\mathrm{val}}\downarrow$ & Obs. NRMSE $\downarrow$ \\
      \midrule
      \multicolumn{4}{l}{\textit{Screen}} \\
      400   & 100 & 0.14791 & 0.09169 \\
      600   & 67  & 0.13173 & 0.06198 \\
      800   & 50  & 0.12064 & 0.04540 \\
      1,200 & 61  & 0.11331 & 0.05758 \\
      1,600 & 71  & \textbf{0.07691} & 0.05824 \\
      2,000 & 63  & 0.08018 & 0.05509 \\
      \addlinespace[2pt]
      \multicolumn{4}{l}{\textit{Confirm}} \\
      40    & 63  & 0.49786 & 0.05453 \\
      100   & 100 & 0.20732 & 0.09105 \\
      200   & 100 & 0.11970 & 0.09202 \\
      $1{,}600^{\dagger}$ & 70 & \textbf{0.05241} & 0.06546 \\
      \bottomrule
    \end{tabular*}
  \end{minipage}}
  \label{tab:appendix_guidance_ns_ks}
\end{table}

\FloatBarrier

The selected configurations were $(N,\zeta_u,\zeta_f)=(400,10,20)$ for Poisson, $(N,\zeta)=(1{,}600,5)$ for NS, and $(N,\zeta)=(1{,}600,70)$ for KS. Increasing the number of Poisson sampling steps from 400 to 800 yielded no further improvement. For NS, increasing the number of steps from 1,600 to 2,000 reduced $J_{\mathrm{val}}$ by only 1.3\% at 25\% greater sampling cost, so the 1,600-step configuration was retained. For KS, the validation score reached its minimum at 1,600 steps. These configurations were used for all reported test evaluations and the inference-cost measurements in
Table~\ref{tab:sampling_cost}.

\FloatBarrier

\section{Ensemble-size sensitivity}
\label{sec:appendix_ensemble_size}

\setcounter{table}{0}
\setcounter{figure}{0}

The common ensemble size was selected using C-EDM validation results. 
The balanced Poisson panel contained 200 fields, with ten cases for each combination of $k_c$ and observation fraction.
The KS panel contained 125 trajectories, with five cases for each combination of viscosity interval and observation fraction. 
The NS panel contained 30 trajectories evaluated at all five observation fractions, yielding 150 reconstruction cases. 
Validation fields or trajectories, masks, and sampling seeds were fixed across ensemble sizes. 
For each case, a single 32-member ensemble was generated, and smaller ensemble sizes were evaluated using nested prefixes of the same members.

\begin{figure}[!t]
  \centering
  \includegraphics[width=0.88\textwidth]{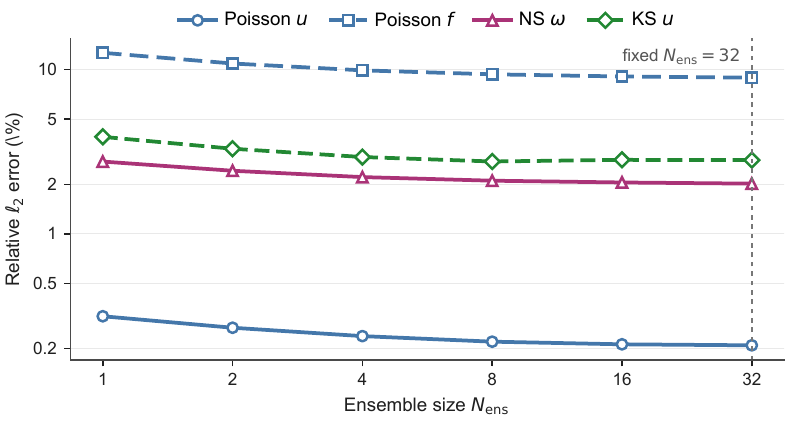}
  \caption{C-EDM ensemble-mean full-field relative $\ell_2$ error versus ensemble size for the four reconstruction targets on the balanced validation panels summarized in Table~\ref{tab:appendix_ensemble_size}.}
  \label{fig:appendix_ensemble_size}
\end{figure}

Table~\ref{tab:appendix_ensemble_size} and Figure~\ref{fig:appendix_ensemble_size} summarize the resulting ensemble-size dependence. For Poisson and NS, ensemble-mean reconstruction error decreased through $N_{\mathrm{ens}}=32$. The relative improvement from 16 to 32 members was only 1.4--1.7\%. For KS, ensemble-mean reconstruction error showed no consistent improvement beyond eight members. The KS uncertainty estimation nevertheless improved from 16 to 32 members, with empirical coverage increasing from 85.6\% to 89.3\% and normalized CRPS decreasing from 0.00778 to 0.00758. We therefore retained $N_{\mathrm{ens}}=32$ as the common reporting size for both EDM formulations despite the diminishing gains in reconstruction accuracy.

\begin{table}[!t]
  \centering
  \small
  \setlength{\tabcolsep}{3.2pt}
  \tbl{C-EDM ensemble-size sensitivity on balanced validation panels. $N_{\mathrm{data}}$ denotes the number of independent fields or trajectories, and $N_{\mathrm{case}}$ denotes the number of reconstruction conditions. For each case, results for smaller ensemble sizes use nested prefixes of the corresponding 32-member ensemble.}{%
  \begin{tabular}{lrrrrrrrr}
    \toprule
    Field & $N_{\mathrm{data}}$ & $N_{\mathrm{case}}$ & $N_{\mathrm{ens}}=1$ & $2$ & $4$ & $8$ & $16$ & $32$ \\
    \midrule
    \multicolumn{9}{l}{\textit{Ensemble-mean full relative $\ell_2$ error (\%; $\downarrow$)}} \\
    Poisson $u$ & 200 & 200 & 0.3149 & 0.2680 & 0.2383 & 0.2204 & 0.2124 & \textbf{0.2095} \\
    Poisson $f$ & 200 & 200 & 12.6736 & 10.9102 & 9.9086 & 9.3714 & 9.0919 & \textbf{8.9416} \\
    NS $\omega$ & 30 & 150 & 2.7567 & 2.4230 & 2.2171 & 2.1078 & 2.0559 & \textbf{2.0227} \\
    KS $u$ & 125 & 125 & 3.8993 & 3.3000 & 2.9383 & \textbf{2.7599} & 2.8239 & 2.8134 \\
    \addlinespace[2pt]
    \multicolumn{9}{l}{\textit{KS distributional metrics over unobserved locations}} \\
    Normalized CRPS $\downarrow$ & 125 & 125 & 0.01653 & 0.01203 & 0.00938 & 0.00778 & 0.00778 & \textbf{0.00758} \\
    90\% coverage (\%) $\to 90$ & 125 & 125 & 0.0 & 32.1 & 60.3 & 77.2 & 85.6 & \textbf{89.3} \\
    \bottomrule
  \end{tabular}}
  \label{tab:appendix_ensemble_size}
\end{table}

\end{document}